\documentclass[twocolumn]{aastex61}
\newcommand{\ha}{H$\alpha$}
\newcommand{\hb}{H$\beta$}

\newcommand{\hei}{He~{\sc i}}
\newcommand{\heii}{He~{\sc ii}}
\newcommand{\fei}{Fe~{\sc i}}
\newcommand{\feii}{Fe~{\sc ii}}
\newcommand{\caii}{Ca~{\sc ii}}
\newcommand{\crii}{Cr~{\sc ii}}
\newcommand{\nai}{Na~{\sc i}}
\newcommand{\oi}{O~{\sc i}}

\newcommand{\nitrogeni}{N~{\sc i}}

\newcommand{\niii}{Ni~{\sc ii}}
\newcommand{\scii}{Sc~{\sc ii}}
\newcommand{\ci}{C~{\sc i}}
\newcommand{\cii}{C~{\sc ii}}
\newcommand{\mgi}{Mg~{\sc i}}
\newcommand{\mgii}{Mg~{\sc ii}}
\newcommand{\tiii}{Ti~{\sc ii}}
\newcommand{\sii}{Si~{\sc i}}
\newcommand{\siii}{Si~{\sc ii}}
\newcommand{\siiii}{Si~{\sc iii}}
\newcommand{\kms}{$\rm km\,s^{-1}$}
\newcommand{\vp}{$v_{\rm p}$}

\newcommand{\vsini}{$v\,{\rm sin}\,i$}

\newcommand{\msun}{$M_{\rm \odot}$}
\newcommand{\rsun}{$R_{\rm \odot}$}
\newcommand{\lsun}{$L_{\rm \odot}$}
\newcommand{\teff}{$T_{\rm eff}$}

\newcommand{\rv}{$V_{r}$}
\newcommand{\vcrit}{$v_{\rm crit}$}
\newcommand{\vrot}{$v_{\rm rot}$}

\shorttitle{HD 55606}
\shortauthors{Chojnowski}

\begin{document}

\title{The Remarkable Be+sdOB Binary HD 55606 I: Orbital and Stellar Parameters\footnote{Based on observations obtained with the Apache Point Observatory 3.5-meter telescope, which is owned and operated by the Astrophysical Research Consortium.}}

\author{S. Drew Chojnowski}
\affiliation{Apache Point Observatory and New Mexico State University, P.O. Box 59, Sunspot, NM, 88349-0059, USA}
\affiliation{Observer, Apache Point Observatory 3.5m Telescope}

\author{Jonathan Labadie-Bartz}
\affiliation{Department of Physics, Lehigh University, Bethlehem, PA 18015, USA}
\affiliation{Department of Physics \& Astronomy, University of Delaware, Newark, DE 19716, USA}

\author{Thomas Rivinius}
\affiliation{ESO – European Organisation for Astronomical Research in the Southern Hemisphere, Casilla 19001, Santiago 19, Chile}

\author{Douglas Gies}
\affiliation{Center for High Angular Resolution Astronomy and Department of Physics and Astronomy, Georgia State University, P. O. Box 5060, Atlanta, GA 30302-5060, USA}
\affiliation{Observer, Apache Point Observatory 3.5m Telescope}

\author{	Despina Panoglou}
\affiliation{Observat\'orio Nacional, Rua General Jos\'e Cristino 77, 20921-400,S\~ao Cristov\~ao, Rio de Janeiro, Brazil}

\author{Marcelo Borges Fernandes}
\affiliation{Observat\'orio Nacional, Rua General Jos\'e Cristino 77, 20921-400,S\~ao Cristov\~ao, Rio de Janeiro, Brazil}

\author{John P. Wisniewski}
\affiliation{Department of Physics \& Astronomy, The University of Oklahoma, 440 W. Brooks St. Norman, OK 73019, USA}
\affiliation{Observer, Apache Point Observatory 3.5m Telescope}

\author{David G. Whelan}
\affiliation{Department of Physics, Austin College, 900 N. Grand Ave., Sherman, TX 75090, USA}

\author{Ronald E. Mennickent}
\affiliation{Departamento de Astronom\'ia, Universidad de Concepci\'on, Concepci\'on, Chile}

\author{Russet McMillan}
\affiliation{Apache Point Observatory and New Mexico State University, P.O. Box 59, Sunspot, NM, 88349-0059, USA}
\affiliation{Observer, Apache Point Observatory 3.5m Telescope}

\author{Jack M. Dembicky}
\affiliation{Apache Point Observatory and New Mexico State University, P.O. Box 59, Sunspot, NM, 88349-0059, USA}
\affiliation{Observer, Apache Point Observatory 3.5m Telescope}

\author{Candace Gray}
\affiliation{Apache Point Observatory and New Mexico State University, P.O. Box 59, Sunspot, NM, 88349-0059, USA}
\affiliation{Observer, Apache Point Observatory 3.5m Telescope}

\author{Ted Rudyk}
\affiliation{Apache Point Observatory and New Mexico State University, P.O. Box 59, Sunspot, NM, 88349-0059, USA}
\affiliation{Observer, Apache Point Observatory 3.5m Telescope}

\author{Guy S. Stringfellow}
\affiliation{Center for Astrophysics and Space Astronomy, Department of Astrophysical and Planetary Sciences, University of Colorado, 389 UCB, Boulder, Colorado 80309-0389, USA}
\affiliation{Observer, Apache Point Observatory 3.5m Telescope}

\author{Kathryn Lester}
\affiliation{Center for High Angular Resolution Astronomy and Department of Physics and Astronomy, Georgia State University, P. O. Box 5060, Atlanta, GA 30302-5060, USA}
\affiliation{Principal Investigator, Apache Point Observatory 3.5m Telescope}

\author{Sten Hasselquist}
\affiliation{Apache Point Observatory and New Mexico State University, P.O. Box 59, Sunspot, NM, 88349-0059, USA}
\affiliation{Observer, Apache Point Observatory 3.5m Telescope}

\author{Sergey Zharikov}
\affiliation{Observatorio Astron\'omico Nacional SPM, Instituto de Astronom\'{\i}a, UNAM, Ensenada, BC, Mexico}

\author{Ronaldo Levenhagen}
\affiliation{Universidade Federal de S\~ao Paulo, Departamento de F\'isica, Rua Prof. Artur Riedel, 275, CEP 09972–270, Diadema, SP, Brazil}

\author{Tiago Souza}
\affiliation{Observat\'orio Nacional, Rua General Jos\'e Cristino 77, 20921-400,S\~ao Cristov\~ao, Rio de Janeiro, Brazil}

\author{Nelson Leister}
\affiliation{Instituto de Astronomia, Geof\'isica e Ci\^encias Atmosf\'ericas, Universidade de S\~ao Paulo, Brazil}

\author{Keivan Stassun}
\affiliation{Department of Physics and Astronomy, Vanderbilt University, Nashville, TN 37235, USA}

\author{Robert J. Siverd}
\affiliation{Las Cumbres Observatory Global Telescope Network, 6740 Cortona Drive, Suite 102, Santa Barbara, CA 93117, USA}

\author{Steven R. Majewski}
\affiliation{Department of Astronomy, University of Virginia, P.O. Box 400325, Charlottesville, VA 22904-4325, USA}

\begin{abstract}
Prompted by peculiar spectroscopic variability observed in SDSS/APOGEE $H$-band spectra, we monitored the Be star HD~55606 using optical spectroscopy and found that it is an exotic double-lined spectroscopic binary (SB2) consisting of a Be star and a hot, compact companion that is probably an OB subdwarf (sdOB) star. Motion of the sdOB star is traced by its impact on the strong {\hei} lines, observed as radial velocity ({\rv}) variable, double-peaked emission profiles with narrow central absorption cores. Weak {\heii}~4686~{\AA} absorption associated with the companion star is detected in most spectra. Use of the emission peaks of low-ionization emission lines to trace the Be star {\rv} and the {\hei} lines to trace the companion star {\rv} yields a circular orbital solution with a 93.8-day period and masses of $M_{\rm Be}=6.2$ {\msun} and $M_{\rm sdOB}=0.9$ {\msun} in the case of $i=80^{\circ}$. HD~55606 exhibits a variety of phase-locked variability, including the development of shell lines twice per orbit. The shell phases coincide with variation in the double emission peak separations, and both forms of variability are likely caused by a two-armed spiral density perturbation in the Be disk. The intensity ratios of the double emission peaks are also phase-locked, possibly indicating heating by the sdOB star of the side of the Be disk facing it. HD 55606 is a new member of the growing sample of Be+sdOB binaries, in which the Be star's rapid rotation and ability to form a disk can be attributed to past mass transfer. 

\end{abstract}


\section{Introduction} \label{intro}
Perhaps the strongest motivation for studying classical Be stars is the fact that the mechanism(s) responsible for their ability/tendency to form circumstellar gas disks remains unknown despite more than a hundred years of research \citep{rivi13a}. It is well known by now that Be stars are the most rapidly-rotating class of non-degenerate stars and that rapid rotation is a requirement for forming a Be disk, and it is suspected that the interactions of multiple non-radial pulsation frequencies may provide the added boost needed for the star to eject their outer layers into orbiting Keplerian disks \citep{rivi98,baade16}. Despite progress in determining how the disks are formed, the question of why the stars are rotating so rapidly in the first place remains unanswered.

In a handful of cases involving very rapidly-rotating Be stars ({\vsini} $>300$ {\kms}), the spin-up of the Be star can be explained via mass and angular momentum transfer with a binary companion. The Be star in these cases is the fast rotating mass gainer, while the companion is a hot, stripped-down core of a formerly massive star. Examples include the bright stars $\phi$~Per \citep{gies98}, FY~CMa \citep{peters08}, 59~Cyg \citep{peters13}, HR~2142 \citep{peters16}, and 60~Cyg \citep{wang17}. In each case, the hot companion was initially suspected due either to radial velocity ({\rv}) variability of {\hei} emission or {\heii}~4686~{\AA} absorption, or in the case of HR~2142, to detection of transient narrow absorption or `shell' phases \citep{peters83}. Thanks to the higher flux ratio ($F_{\rm sdO}/F_{\rm Be}$) in the ultraviolet (UV) as compared to the optical, the photospheric signatures of the sdO companions of the aforementioned Be stars have all been directly detected via UV spectroscopy. $o$~Pup \citep{koubsky12b} and HD~161306 \citep{koubsky14} are likely additional examples of Be+sdO binaries, but UV spectra of these systems have not been investigated to date.

Recently, \citet{wang18} more than doubled the sample of known and suspected Be+sdO binaries by cross-correlating archival UV spectra of classical Be stars with sdO star model spectra. An sdO star signature was found in 12 out of 264 stars studied, and for 8 of these, {\rv} variability was detected in the UV spectra. The authors of this work suggest that the fraction of Be stars spun up through binary interaction may be significantly higher than previously thought, with detection of the sdO star being impossible depending on evolutionary status, and orbital motion of the Be stars virtually impossible to detect without a large collection of high-resolution spectra. Indeed, measurement of the orbital motions of the Be stars in the known and suspected Be+sdO systems typically relies on indirect measurements of Be disk emission lines, and the resulting {\rv} semi-amplitudes are typically quite small ($K_{\rm 1}\sim10$ {\kms}). 

Here we present strong evidence that the star HD~55606 ($V\sim9.4$ mag.) is another member of the growing class of Be+sdO binaries, and may in fact be the first known Be+sdB binary. The initial clues were provided by high-resolution, $H$-band spectroscopy of the star by the Apache Point Observatory Galactic Evolution Experiment \citep[APOGEE,][]{majewski16}, one of the sub-surveys of the third installment of the Sloan Digital Sky Survey \citep[SDSS-III,][]{eisenstein2011}. For the vast majority of the classical Be stars, the only lines detected in APOGEE spectra are from the Brackett series of hydrogen \citep{choj15}, but the spectra of HD~55606 (referred to as ABE-A15 in that paper) were found to contain numerous double-peaked emission lines from neutral metallic species including {\ci}, {\mgi}, and {\fei}, with the {\ci}~16895~{\AA} line being particularly strong. \citet{choj17} used the Brackett series emission peaks of HD~55606 to measure {\rv} from the APOGEE spectra, and found them to vary by $\sim10$ {\kms} in three spectra covering 26 days. Remarkably, the double emission peak velocity separations ({\vp}) of most lines increased by $\sim$60 {\kms} in the 21 days between the first and second spectrum. The unique nature of these peculiarities led us to suspect an external influence on the disk, i.e. that perhaps HD~55606 was an interacting binary. 

Our follow-up optical spectroscopy campaign has subsequently revealed numerous clues suggesting HD~55606 is indeed an interacting binary, beginning with detection of narrow {\hei} emission/absorption components and of weak {\heii}~4686~{\AA} absorption that move in anti-phase relative to the Be disk. As with the previously known Be+sdO binaries, HD~55606 also exhibits a variety of phase-locked variability including reversal of the double peak intensity ratios once per orbit and variable emission peak separations that return to the average value twice per orbit. In addition, HD~55606 is a transient Be shell star, with just over half of the available spectra exhibiting narrow, blueshifted (with respect to the Be star) absorption components in the Balmer series lines, the {\oi}~7771-7775~{\AA} triplet, and occasionally in numerous additional lines.

Section~\ref{obs} of this paper provides a description of the currently available HD~55606 data, including high-resolution optical and $H$-band spectroscopy. In Section~\ref{basicspectrum} we describe the basic properties of the spectra, and in Sections~\ref{rvs} and \ref{orbit}, the associated radial velocity measurements and resulting orbital parameters are presented. Basic stellar parameters of the binary components are estimated in Section~\ref{stepars}, while Section~\ref{phaselocked} describes the orbital-phase-locked variability observed in the system. In Section~\ref{discussion}, disk and Roche lobe radii are estimated and a model helping to explain the phase-locked variability is presented. The findings are summarized in Section~\ref{conclusions}.

\begin{deluxetable*}{rccccrrcrr}
\tablecaption{Spectroscopic Data Summary \label{specobstable}}
\tabletypesize{\scriptsize}
\tablehead{ 
\colhead{\#} & \colhead{Telescope/Instrument} & \colhead{Date} & \colhead{HJD} & \colhead{$t_{exp}$} & \colhead{Airmass} & \colhead{$R$}                       & \colhead{Range}   & \colhead{SNR}  & \colhead{SNR} \\
             &                                &                &               & \colhead{[sec]}     &                   & \colhead{($\lambda/\Delta\lambda$)} & \colhead{[{\AA}]} & \colhead{Blue} & \colhead{Red}
}
\startdata
1 & ESO 1.52m/FEROS & 2001-10-09 & 2452191.90433 & 400 & 1.4 & 48,000 & 3600--9200 & 85 & 83 \\
2 & OHP 1.93m/ELODIE (BeSS) & 2001-12-22 & 2452265.53776 & 5402 & 1.5 & 42,000 & 3900--6800 & 18 & 49 \\
3 & Sloan 2.5m/APOGEE & 2013-12-19 & 2456645.88162 & 2002 & 1.2 & 22,500 & 15145--16960 & \nodata & 213 \\
4 & Sloan 2.5m/APOGEE & 2014-01-09 & 2456666.78579 & 3003 & 1.2 & 22,500 & 15145--16960 & \nodata & 249 \\
5 & Sloan 2.5m/APOGEE & 2014-01-14 & 2456671.77185 & 2002 & 1.2 & 22,500 & 15145--16960 & \nodata & 198 \\
6 & APO 3.5m/ARCES & 2014-01-14 & 2456671.77958 & 720 & 1.2 & 31,500 & \phantom{0}3500--10200 & 127 & 168 \\
7 & APO 3.5m/ARCES & 2016-03-26 & 2457473.58413 & 1200 & 1.2 & 31,500 & \phantom{0}3500--10200 & 114 & 183 \\
8 & ESO 2.2m/FEROS & 2016-04-13 & 2457491.79645 & 650 & 1.6 & 48,000 & 3600--9200 & 82 & 80 \\
9 & APO 3.5m/ARCES & 2016-12-24 & 2457746.79520 & 1000 & 1.3 & 31,500 & \phantom{0}3500--10200 & 118 & 143 \\
10 & APO 3.5m/ARCES & 2017-02-08 & 2457792.86471 & 1200 & 2.0 & 31,500 & \phantom{0}3500--10200 & 87 & 115 \\
11 & APO 3.5m/ARCES & 2017-02-11 & 2457795.72327 & 1200 & 1.2 & 31,500 & \phantom{0}3500--10200 & 118 & 151 \\
12 & APO 3.5m/ARCES & 2017-02-12 & 2457796.68814 & 1200 & 1.2 & 31,500 & \phantom{0}3500--10200 & 117 & 158 \\
13 & APO 3.5m/ARCES & 2017-03-09 & 2457821.67007 & 1200 & 1.2 & 31,500 & \phantom{0}3500--10200 & 136 & 150 \\
14 & APO 3.5m/ARCES & 2017-03-12 & 2457824.74712 & 1200 & 1.6 & 31,500 & \phantom{0}3500--10200 & 83 & 130 \\
15 & APO 3.5m/ARCES & 2017-04-06 & 2457849.62973 & 1500 & 1.3 & 31,500 & \phantom{0}3500--10200 & 139 & 177 \\
16 & SPM 2.1m/RFOSC & 2017-10-27 & 2458053.96493 & 2700 & 1.3 & 18,000 & 3770--7600 & 127 & 161 \\
17 & APO 3.5m/ARCES & 2017-11-10 & 2458067.84857 & 1200 & 1.7 & 31,500 & \phantom{0}3500--10200 & 106 & 180 \\
18 & APO 3.5m/ARCES & 2017-11-12 & 2458069.82557 & 1200 & 1.9 & 31,500 & \phantom{0}3500--10200 & 62 & 146 \\
19 & APO 3.5m/ARCES & 2017-11-26 & 2458083.84974 & 1200 & 1.4 & 31,500 & \phantom{0}3500--10200 & 111 & 207 \\
20 & APO 3.5m/ARCES & 2017-11-29 & 2458086.81632 & 1200 & 1.5 & 31,500 & \phantom{0}3500--10200 & 100 & 174 \\
21 & APO 3.5m/ARCES & 2017-12-02 & 2458089.78814 & 1320 & 1.7 & 31,500 & \phantom{0}3500--10200 & 89 & 163 \\
22 & APO 3.5m/ARCES & 2017-12-04 & 2458091.81341 & 1200 & 1.4 & 31,500 & \phantom{0}3500--10200 & 110 & 170 \\
23 & SPM 2.1m/RFOSC & 2017-12-07 & 2458095.03748 & 1200 & 1.7 & 18,000 & 3770--7600 & 65 & 87 \\
24 & APO 3.5m/ARCES & 2017-12-08 & 2458095.81522 & 1200 & 1.4 & 31,500 & \phantom{0}3500--10200 & 123 & 162 \\
25 & APO 3.5m/ARCES & 2017-12-09 & 2458097.05647 & 1200 & 2.6 & 31,500 & \phantom{0}3500--10200 & 22 & 123 \\
26 & APO 3.5m/ARCES & 2017-12-10 & 2458097.79188 & 1200 & 1.5 & 31,500 & \phantom{0}3500--10200 & 125 & 177 \\
27 & APO 3.5m/ARCES & 2017-12-11 & 2458098.82613 & 1200 & 1.3 & 31,500 & \phantom{0}3500--10200 & 114 & 179 \\
28 & APO 3.5m/ARCES & 2017-12-12 & 2458099.80512 & 1900 & 1.4 & 31,500 & \phantom{0}3500--10200 & 115 & 191 \\
29 & APO 3.5m/ARCES & 2017-12-13 & 2458100.81465 & 1200 & 1.3 & 31,500 & \phantom{0}3500--10200 & 127 & 215 \\
30 & APO 3.5m/ARCES & 2017-12-14 & 2458101.80575 & 1800 & 1.3 & 31,500 & \phantom{0}3500--10200 & 170 & 246 \\
31 & APO 3.5m/ARCES & 2017-12-24 & 2458111.81305 & 2000 & 1.3 & 31,500 & \phantom{0}3500--10200 & 135 & 154 \\
32 & APO 3.5m/ARCES & 2017-12-25 & 2458112.86499 & 1200 & 1.2 & 31,500 & \phantom{0}3500--10200 & 115 & 183 \\
33 & APO 3.5m/ARCES & 2017-12-26 & 2458113.81601 & 1200 & 1.2 & 31,500 & \phantom{0}3500--10200 & 147 & 211 \\
34 & APO 3.5m/ARCES & 2017-12-29 & 2458116.81186 & 1500 & 1.2 & 31,500 & \phantom{0}3500--10200 & 157 & 192 \\
35 & APO 3.5m/ARCES & 2017-12-30 & 2458117.81787 & 1500 & 1.2 & 31,500 & \phantom{0}3500--10200 & 131 & 193 \\
36 & APO 3.5m/ARCES & 2017-12-31 & 2458118.79662 & 1500 & 1.3 & 31,500 & \phantom{0}3500--10200 & 152 & 216 \\
37 & APO 3.5m/ARCES & 2018-01-01 & 2458119.79181 & 1200 & 1.3 & 31,500 & \phantom{0}3500--10200 & 161 & 196 \\
38 & APO 3.5m/ARCES & 2018-01-02 & 2458120.73963 & 2100 & 1.4 & 31,500 & \phantom{0}3500--10200 & 112 & 196 \\
39 & APO 3.5m/ARCES & 2018-01-05 & 2458123.78072 & 2400 & 1.3 & 31,500 & \phantom{0}3500--10200 & 158 & 195 \\
40 & APO 3.5m/ARCES & 2018-01-12 & 2458130.81292 & 2400 & 1.2 & 31,500 & \phantom{0}3500--10200 & 172 & 221 \\
41 & APO 3.5m/ARCES & 2018-01-16 & 2458134.81332 & 1500 & 1.2 & 31,500 & \phantom{0}3500--10200 & 97 & 138 \\
42 & APO 3.5m/ARCES & 2018-01-19 & 2458137.79604 & 1200 & 1.2 & 31,500 & \phantom{0}3500--10200 & 125 & 226 \\
43 & APO 3.5m/ARCES & 2018-01-26 & 2458144.79619 & 1200 & 1.3 & 31,500 & \phantom{0}3500--10200 & 128 & 191 \\
44 & APO 3.5m/ARCES & 2018-01-30 & 2458148.75117 & 1200 & 1.2 & 31,500 & \phantom{0}3500--10200 & 138 & 199 \\
45 & APO 3.5m/ARCES & 2018-02-01 & 2458150.75011 & 1500 & 1.2 & 31,500 & \phantom{0}3500--10200 & 140 & 166 \\
46 & APO 3.5m/ARCES & 2018-02-02 & 2458151.79571 & 1800 & 1.3 & 31,500 & \phantom{0}3500--10200 & 156 & 195 \\
47 & APO 3.5m/ARCES & 2018-02-05 & 2458154.79267 & 900 & 1.3 & 31,500 & \phantom{0}3500--10200 & 138 & 206 \\
48 & APO 3.5m/ARCES & 2018-02-08 & 2458157.79755 & 1500 & 1.4 & 31,500 & \phantom{0}3500--10200 & 145 & 197 \\
49 & APO 3.5m/ARCES & 2018-02-22 & 2458171.76730 & 1200 & 1.4 & 31,500 & \phantom{0}3500--10200 & 124 & 181 \\
50 & APO 3.5m/ARCES & 2018-02-25 & 2458174.79054 & 1800 & 1.6 & 31,500 & \phantom{0}3500--10200 & 126 & 159 \\
51 & APO 3.5m/ARCES & 2018-02-27 & 2458176.79054 & 1200 & 1.2 & 31,500 & \phantom{0}3500--10200 & 149 & 200 \\
52 & APO 3.5m/ARCES & 2018-03-02 & 2458179.76934 & 1200 & 1.6 & 31,500 & \phantom{0}3500--10200 & 96 & 162 \\
\enddata
\tablecomments{The HJD column refers to the exposure start times, and the signal-to-noise ratio (SNR) columns pertain to averages of multiple measurements of continuum around 4200~{\AA} (SNR Blue), 6600~{\AA} (SNR Red, optical), or 16650~{\AA} (SNR Red, $H$-band).}
\end{deluxetable*}

\section{Data} \label{obs}

%\subsection{Optical and $H$-band Spectroscopy} \label{opticalspec}
This work makes uses of a total of 52 spectra, including numerous high-resolution optical spectra and three high-resolution $H$-band spectra. The data are summarized in Table~\ref{specobstable}, with the ``SNR Blue'' column giving the signal-to-noise ratio (SNR) around 4200~{\AA} and the ``SNR Red'' column giving the SNR around 6600~{\AA} (optical data) or 16650~{\AA} ($H$-band data). For all spectra used in this paper, the continuum flux was set to unity by fitting spline functions to regions surrounding spectral lines using the {\it continuum} procedure of the Image Reduction and Analysis Facility (IRAF\footnote{IRAF is distributed by the National Optical Astronomy Observatories, which are operated by the Association of Universities for Research in Astronomy, Inc., under cooperative agreement with the National Science Foundation.}). For the echelle spectra, the orders were trimmed so as to allow a small overlap between adjacent orders, paying careful attention to overlapping lines used for measurements. Finally, the orders from each observation were averaged to make single one-dimensional spectra. Heliocentric corrections were applied to all spectra. Here we list the instrument parameters and data reduction methods.

\textbf{APOGEE:} HD 55606 was observed three times by the APOGEE instrument while connected to the Sloan Foundation 2.5m telescope \citep{gunn06} at Apache Point Observatory (APO). APOGEE is a 300-fiber, high-resolution ($R\sim22,500$) NIR spectrograph capturing most of the $H$-band, with two small gaps of $\sim50$~{\AA} and $\sim30$~{\AA}. The details of APOGEE data reduction are described in \citet{nidever15}, and \citet{choj15} provides a general description of the $H$-band spectra of Be stars. The signal-to-noise ratios (SNR) of the APOGEE spectra are quite high (SNR$>$250) due to the 30--50 minute total exposure times.

\textbf{ARCES:} We obtained numerous optical spectra of HD~55606 using the APO 3.5m telescope and the Astrophysical Research Consortium Echelle Spectrograph \citep[ARCES,][]{wang03}. In each exposure, ARCES covers the full optical spectrum (3500--10,000 {\AA}) at a resolution of $R\sim31,500$, recording the light in 107 orders on a 2048x2048 SITe CCD. We used standard IRAF echelle data reduction techniques, including bias subtraction, scattered light and cosmic ray removal, flat-field correction, and wavelength calibration via Thorium-Argon lamp exposures.  

\textbf{FEROS:} We also used two high-resolution optical spectra obtained previously by Ronaldo Levenhagen \& Nelson Leister (2001 October 9) and Marcelo Borges Fernandes (2016 April 13) from the FEROS spectrograph \citep{kaufer99}, while attached to either the European Southern Observatory (ESO) 1.52m telescope (2001 spectrum) or the Max Planck Gesellschaft MPG-ESO 2.2m telecope (2016 spectrum). Both telescopes are at La Silla Observatory. Fiber-fed Extended Range Optical Spectrograph (FEROS) is a bench-mounted Echelle spectrograph with two fibers, each covering a sky area of 2 arcsecond diameter. The instrumental configuration provides a resolution of 0.03 {\AA} / pixel ($R=48000$) in a spectral range 3600--9200~{\AA}. We adopted its complete automatic online reduction pipeline.

\textbf{BeSS/ELODIE:} We downloaded a high-resolution spectrum of HD~55606 from the Be Star Spectra Database \citep[BeSS,][]{neiner11}, which is a forum for both amateur and professional astronomers to archive multi-epoch spectra of Be stars. The spectrum was recorded by the ELODIE echelle spectrograph \citep[$R=42,000$, 3850--6800 {\AA},][]{baranne96} attached to the 1.93m telescope at Haute-Provence Observatory. The SNR of the ELODIE spectrum declines rapidly toward the blue, such that for example the {\hei} 4471 {\AA} line is not discernible. We restricted our analysis of this spectrum to $\lambda>5800$ {\AA}.

\textbf{SPM:} 
We also obtained two optical spectra of HD~55606 using the RFOSC Echelle Spectrograph \citep{Levine} attached to the 2.12m telescope of Observatorio Astron\'{o}mico Nacional at San Pedro M\'{a}rtir (OAN SPM), Mexico. This spectrograph provides spectra over 29 orders, covering the spectral range of 3770---7600 \AA\ with a spectral resolution power of $R\simeq18.000$ at $5000$ \AA, which results in a resolution of $\sim 17$ \kms\ with 2 pixels. The E2V (Marconi) $2048\times 2048$ CCD with a pixel size of $13.5$ $\mu$m was used. A Th---Ar lamp was used for wavelength calibration. All data were reduced by a standard way using the IRAF tasks {\it ccdred} and {\it echelle}. %The calibration lamp was observed at the object position before each scientific exposure. The last allows us to exclude any additional errors in wavelength calibration that was verified repeatedly in different observational runs by observing of radial velocity standard stars using similar instrument setup.  %The wavelength correction of spectra for the earth translational motion was applied using the task {\it rvcorrect}.  

\begin{figure*}
\epsscale{1.0}
\plotone{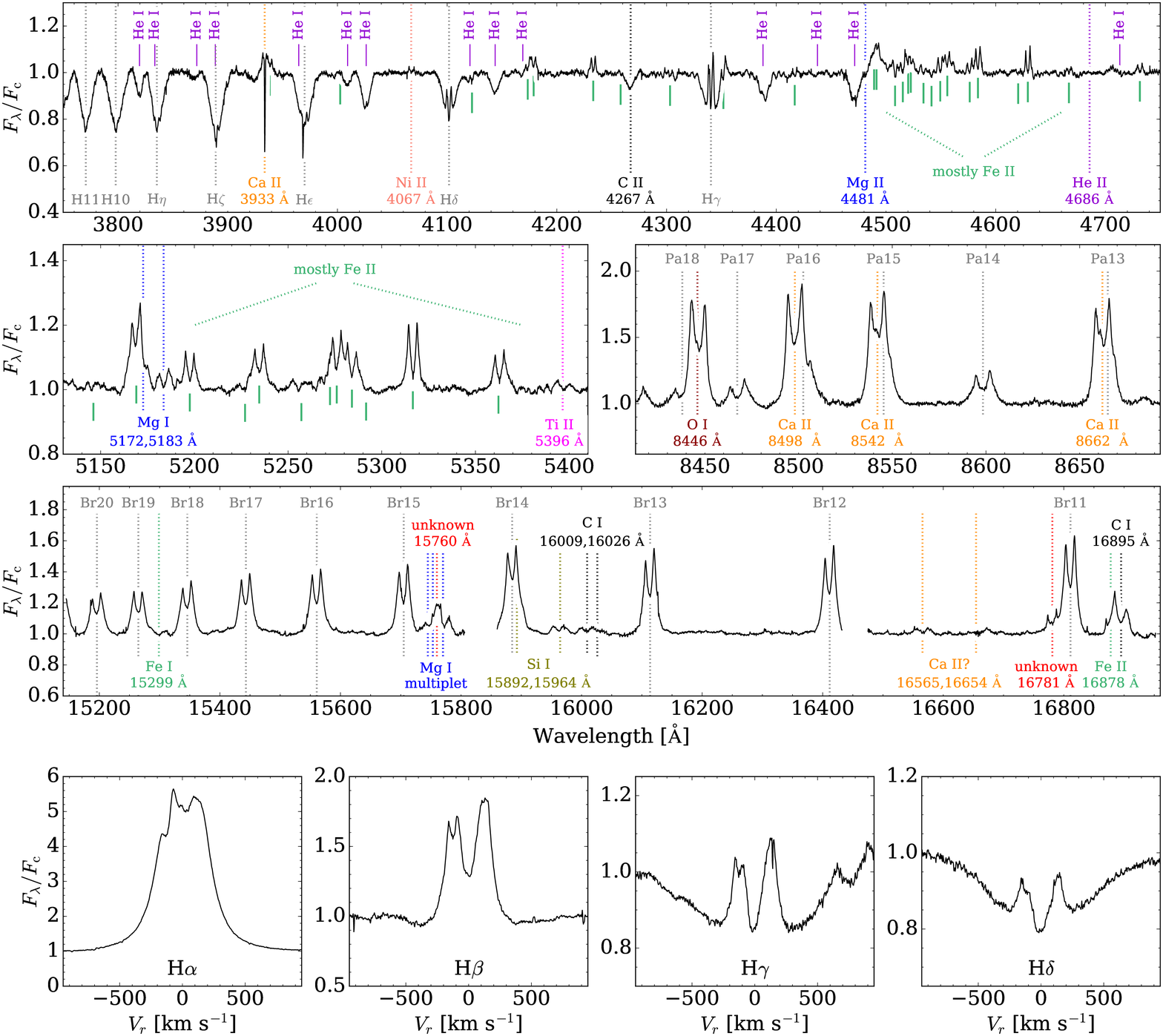}
\caption{Relevant portions of the spectrum of HD 55606 from MJD 56671, where the star was observed simultaneously by both APOGEE and ARCES. The majority of spectral features in the upper rows are labeled, with green ticks below continuum level in the top two rows indicating {\feii} lines. The two gaps in the APOGEE spectrum (third row) are caused by the non-overlapping wavelength coverage of the three APOGEE detectors. The bottom row shows the lower members of the Balmer series on a velocity scale. \label{optical56671}}
\end{figure*}

\section{Spectral Description} \label{basicspectrum}
Figure~\ref{optical56671} displays portions of an ARCES spectrum ($\lambda<10,000$ {\AA}) as well as a full APOGEE spectrum ($\lambda>15,000$ {\AA}) of HD~55606, with both spectra having been serendipitously recorded on the same night (MJD 56671). Here we describe some of the important aspects of the spectra. 

\subsection{Spectral Type} \label{spectype}
The spectral type and stellar parameters of HD~55606 have been investigated previously by several authors beginning with \citet{merrill42}, who acknowledged the broadness of the stellar absorption lines by including an ``n'' (nebulous) in the assigned B3ne spectral type. The spectral type was later revised by \citet{morgan55} and \citet{hiltner56} who agreed independently on a B1:V:pnne spectral type, with the colons indicating ambiguity in the temperature and luminosity class, and the `p' indicating spectral peculiarities. \citet{houk99} leaned toward HD~55606 being a slightly cooler star at B2/3Vnne, while adding the accurate description ``H$\beta$ strongly in emission; H$\gamma$ core in emission; other lines very broad and washed out.'' More sophisticated analysis of the HD~55606 stellar parameters was done by \citet{levenhagen06} and \citet{fremat06}, with both groups ultimately quoting a B0.5V spectral type (the former including the ``nnpe'' qualifier) and {\teff}$>25$ kK, and $M_{*}>10$ {\msun}. 

The large scatter in literature spectral types is likely due to a combination of factors including the rapid rotation of the Be star, the general lack of useful absorption lines, contamination by emission of the few absorption lines that are present, and emission in the {\hei} lines which upon cursory inspection suggests a very hot Be star. However, the strong {\hei} and {\cii}~4267~{\AA} absorption lines in the HD~55606 spectra, non-detection of absorption from the Be star in higher-ionization lines like {\siiii} and {\heii}, and the fact that the {\hei} emission is caused by the binary companion rather than the Be star all add up to suggest a temperature class of roughly B2. Unfortunately, the spectral-typing-useful {\mgii}~4481~{\AA} line seems to be partially or fully in emission in the spectra, such that it does not help constrain the temperature class.

\subsection{Metallic Emission}
Compared to the majority of Be stars, HD~55606 can be considered extreme in terms of metallic emission line content. In fact, part of our original motivation for obtaining optical spectra of HD 55606 was to look for possible causes of or correlations with the abnormally strong {\ci} emission in the APOGEE $H$-band spectra. Among the existing sample of $>300$ APOGEE-observed classical Be stars, HD~55606 has the second strongest {\ci}~16895~{\AA} emission, with typical equivalent $W_{\rm \lambda}\sim-5$~{\AA}. Along with the {\ci}~16895~{\AA} line, numerous other not-usually-detected (in Be star spectra) metallic emission species are present in the $H$-band spectra of HD~55606, including {\feii}, {\fei}, {\mgi}, {\sii}, and the unidentified lines at 15760~{\AA} and 16781~{\AA} \citep[see][for a discussion of these unidentified lines]{choj15}.

The situation is similar in the optical, with HD~55606 exhibiting an exceptionally rich metallic emission spectrum during all epochs. No forbidden lines are detected in emission. The permitted emission species include {\feii} ($>90$ lines), {\tiii} ($>30$ lines), {\crii} ($>15$ lines), {\scii} ($>10$ lines), {\siii} (4 lines), {\mgi} (5 lines), {\nai} (2 lines), {\oi} (4 lines), {\nitrogeni} (9 lines), {\niii} (4067~{\AA}), and {\caii} (7 lines). The overall strongest metallic emission features in the HD~55606 spectra are the IR {\caii} triplet (8498, 8542, 8662 {\AA}) lines, with e.g. {\caii}~8542~{\AA} having an average height of 1.85 continuum units, and an equivalent width of $\sim7.3$ {\AA} after correcting for the estimated contribution of the blended Paschen series line (i.e. by removing the average of the Pa14 and Pa16 profiles from the {\caii}~8542~{\AA}+Pa15 blend). The {\oi}~8446~{\AA} line is the second strongest metallic emission feature after the {\caii} triplet lines. 

The metallic emission lines of HD~55606 are always purely double-peaked with average velocity separations of 230--240 {\kms}, but the majority of lines exhibit variable peak separations ({\vp}) and intensity ratios of the violet (V) over the red (R) emission peaks (i.e. V/R ratios). {\oi}~8446~{\AA} is the most variable metallic line overall, with {\vp} ranging from 150--250 {\kms} and the V/R ratio varying by $\sim0.15$ continuum units. The enhanced variability of that line may be due to the different formation mechanism, i.e. Ly-$\beta$ fluorescence \citep{mathew12}. The equivalent widths of all metallic lines are roughly constant during all observed epochs however, indicating the Be disk is being fueled at a more or less constant rate.

Some of the neutral metallic emission lines (namely {\mgi} and {\ci}) seem largely immune to the variability evident in other lines, maintaining roughly constant {\vp} that are between $\sim$50--100 {\kms} wider than those of the singly-ionized metals. The larger peak separations indicate these lines have an inner disc origin, and the mere detection of them suggests that the HD~55606 Be disk is particularly dense.

\subsection{Paschen and Brackett Series Emission}
The morphology and temporal behavior of the Paschen and Brackett series emission lines is essentially identical to that of the metallic lines. The line profiles are always double-peaked with variable peak separations and V/R ratios and more or less constant equivalent widths and heights above continuum level. 

\subsection{Balmer Series Emission}
\citet{panoglou18} recently showed that non-standard (triple-peaked, flat-topped) line profiles can result from perturbations caused by the binary companion, and indeed that seems to be the case for HD~55606. Unlike the metallic, Paschen series, and Brackett series lines, the {\ha} line profiles typically exhibit a deformed morphology consisting of $>2$ peaks, with a central emission peak being the dominant one during certain orbital phases. A standard, double-peaked {\ha} line profile is never observed. From {\hb} toward the Balmer break however, the Balmer series assume a progressively more purely-double-peaked morphology, though with the violet peak of the {\hb} and H$\gamma$ line profiles being split in most spectra (see the lowermost row of Figure~\ref{optical56671}). On average, the dominant peaks of H$\beta$, H$\gamma$, and H$\delta$ are separated by 230, 239, and 263~{\kms} respectively. 

\begin{figure}[h!]
\epsscale{1.0}
\plotone{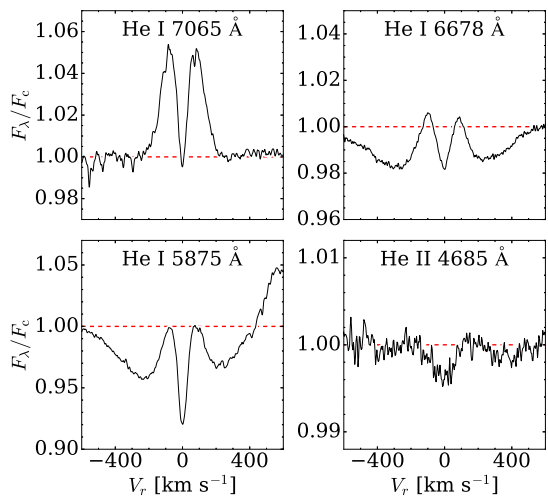}
\caption{Detection of the sdOB star in the redmost {\hei} lines covered and in {\heii}~4686~{\AA}, as seen in an average spectrum in the rest frame of the sdOB star. Dashed red lines indicate continuum level. The feature redward of {\hei}~5876~{\AA} is emission in the sodium D doublet. \label{heidemo}}
\end{figure}

\subsection{{\hei} and {\heii} Features} \label{helium}
The most peculiar aspect of the optical spectra of HD~55606 is the composite nature of the strong {\hei} lines, namely {\hei} 3889, 4026, 4471, 4713, 5015, 5876, 6678, and 7065~{\AA}. Superimposed on the broad photospheric absorption of the Be star are narrow emission + absorption components that migrate in radial velocity across the broad absorption profile from epoch-to-epoch in anti-phase with respect to lines formed in the Be disk. For the blue {\hei} lines, all that is typically seen is a narrow absorption spike embedded in the broad Be star absorption lines, but the {\hei} 5876, 6678, and 7065~{\AA} lines always exhibit a double-peaked emission morphology indicative of formation in a circumstellar disk around the companion star. 

A weak {\heii}~4686~{\AA} absorption line associated with the companion star is present in most spectra. In an average spectrum in the companion star rest frame, as shown in Figure~\ref{heidemo} along with the strong {\hei} lines, the {\heii} line extends just $\sim0.4$\% below continuum level, with an equivalent width of $\sim9$ m{\AA} and a Gaussian full width at half maximum (FWHM) of $\sim160$ {\kms}. The FWHM of the central absorption features in the {\hei} lines on the other hand are significantly smaller at $\sim45$--60~{\kms} such they may be formed primarily by self-absorption of gas in the companion star's accretion disk rather than on the companion star's photosphere. 

\subsection{Transient Shell Phases}
Another unusual aspect of the HD~55606 spectra is the detection in most of the optical spectra of narrow (typical FWHM of $\sim42$ {\kms}) absorption features, or `shell' lines, in the Balmer series lines and in the {\oi}~7771-7775~{\AA} triplet. For the bluemost Balmer series lines, the transient shell lines can extend the depths of the normal/broad Balmer series absorption downward by more than 40\% of continuum level. On several epochs (MJDs 58150--58157), the shell features were detected in a number of additional lines beyond the Balmer series and {\oi} triplet, including the Paschen series, {\siii} (3856, 4128, 4131, 6347, \& 6371~{\AA}), {\feii} (4924, 5018, 5169, \& 5317~{\AA}), and {\niii} (3769, 3849, 4067, and 4187~{\AA}). %In terms of line content, the gas responsible for the shell lines therefore mimics the atmosphere of an early-A supergiant ($T\sim9$--10 kK).

Shell lines in the Balmer series are common for apparently non-binary Be stars viewed edge-on or nearly so, where they are formed due to projection of cool, inner disk gas against the photosphere of the central B star \citep{hanuschik96,rivi06}. In those cases however, the features are permanent for as long the disk is present. {\feii} and {\tiii} are often the among the strong shell lines in spectra of edge-on Be stars. In the case of HD~55606 however, the shell features are transient, appearing in just over half of the spectra. Peculiarly, {\feii} and {\tiii} are typically immune to the shell phases despite HD~55606 having strong {\feii} and {\tiii} emission spectra. Only at maximum shell strength, as judged by Balmer series shell depths or equivalent widths, do we see any hint of {\feii} shell components.

\begin{figure}[ht!]
\epsscale{1.0}
\plotone{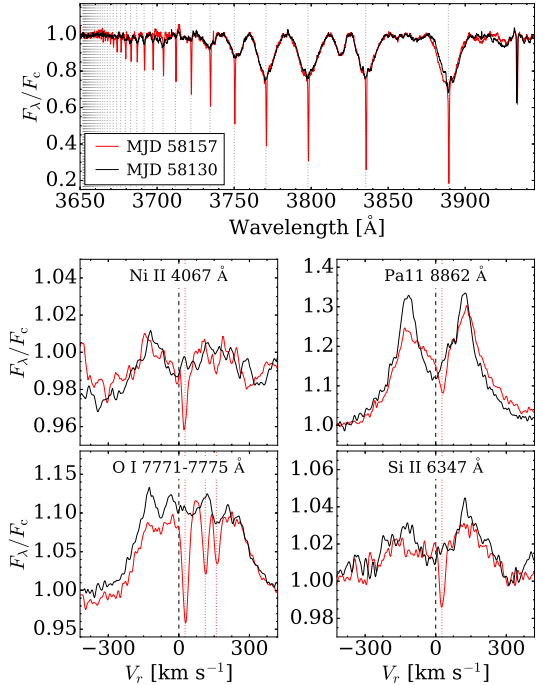}
\caption{Demonstration of a shell epoch (MJD 58157) versus a non-shell epoch (MJD 58130). The top row shows the dramatic effect of the shell phases on the blue Balmer series lines, while the lower panels give a detailed view of the behavior of {\niii}~4067~{\AA}, Pa11, the {\oi}~7771--7775~{\AA} triplet (centered on {\oi}~7771~{\AA}), and {\siii}~6347~{\AA}. In the lower panels, vertical dashed line indicate the Be star rest frame and vertical dotted lines the relative shell velocity. \label{balmershellfig}}
\end{figure}

Figure~\ref{balmershellfig} demonstrates the difference between shell versus non-shell epochs. The MJD 58157 spectrum is just one of four available examples in which shell features were detected in lines beyond the Balmer series and {\oi} triplet. Transient Balmer series shell lines have been reported in the cases of the confirmed Be+sdO binaries HR~2142 \citep{peters72} and FY CMa \citep{rivi04}, but HD~55606 represents the first example of detection of corresponding shell features in optical metal lines.

\section{Radial Velocities} \label{rvs}
\subsection{Be Disk {\rv}}
\begin{figure}[h!]
\epsscale{1.0}
\plotone{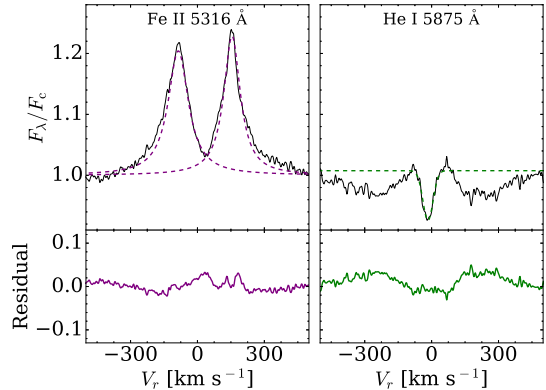}
\caption{An example of Guassian fitting of the {\feii}~5317~{\AA} and {\hei}~5876~{\AA} for radial velocity measurement of the Be and sdOB disk respectively, using the MJD 57849 ARCES spectrum as an example. \label{linefit}}
\end{figure}

Due to the complicated nature of the Be star spectrum, including a high degree of rotational broadening in the few detected absorption lines, contamination throughout the spectrum from weak metallic emission, and the companion's contributions in the {\hei} lines, we used the sharp double peaks of the Paschen series, Brackett series, {\feii}, and {\caii} triplet emission lines to trace the radial velocity ({\rv}) of the Be disk. This was done by interactively fitting Gaussians to the peaks of each line and taking the average as the observed wavelength. Assuming an axisymmetric disk around the Be star, the disk {\rv} should correspond exactly to the stellar {\rv}. Since the measurements are indirect however, we refer to them as `Be disk {\rv}.' The left panels of Figure~\ref{linefit} show an example of fitting of the {\feii}~5317~{\AA} emission peaks. 

From the emission peak measurements we also derived velocity separations and V/R ratios, where the V and R peak heights were measured continuum units. 

\subsection{sdOB Disk and Transient Shell {\rv}}
The same procedure of fitting the double emission peaks was repeated for the {\hei} emission in order to measure the {\rv} of the companion object, but we found far less scatter from Gaussian fitting of the central narrow absorption component. The latter measurements were adopted, permitting use of the blue {\hei} lines (e.g. 3888, 4471, 4713 {\AA}), where the emission peaks are not clearly detected but where the narrow central absorption feature usually is. Up to six {\hei} lines per optical spectrum were used to estimate the companion object {\rv}. Given the complete lack of detected {\hei} lines in the APOGEE's coverage of the $H$-band ({\hei} 1.7 $\mu$m is not covered), the companion {\rv} could not be measured from the APOGEE spectra. The right-hand panels of Figure~\ref{linefit} show an example of Gaussian fitting of the {\hei}~5876~{\AA} narrow absorption component.

The {\rv} of the transient shell components were measured by fitting single Gaussians to the line profiles. Tight agreement of the shell {\rv}, regardless of species, indicates a common origin of the features.

\begin{figure}[t!]
\epsscale{1.0}
\plotone{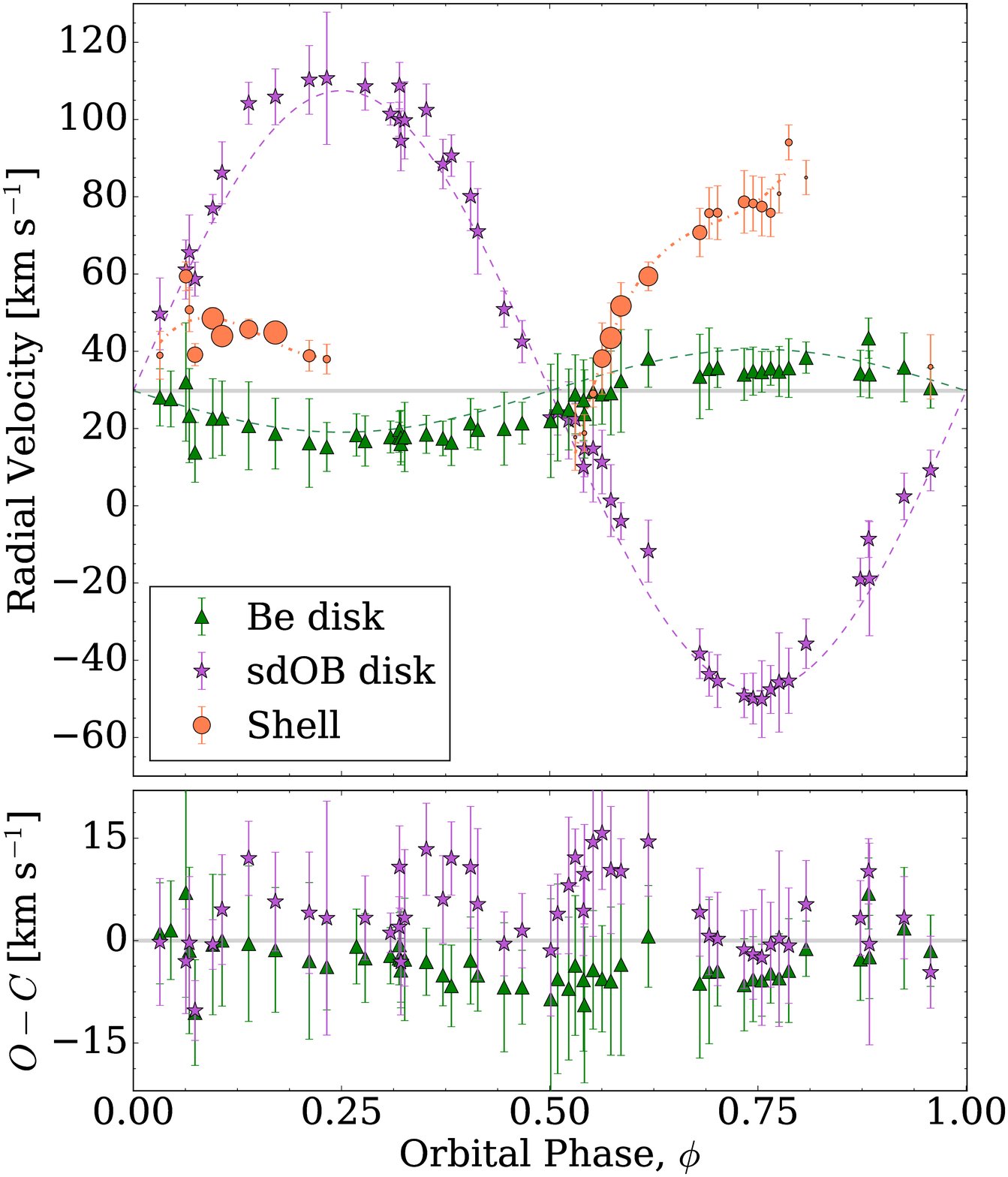}
\caption{Phased {\rv} curves of the Be disk, sdOB disk, and transient shell components, with associated residuals (observed - curve) shown in the lower panel. The sizing of the shell {\rv} points is proportional to the strength of the shell features during a given epoch, as judged by the equivalent width of the H$\eta$ shell absorption. The dashed-dotted lines (orange) are separate third order polynomial fits of the shell {\rv} during the two shell phases.  \label{orbitfig}}
\end{figure}

\begin{deluxetable}{lr}
\tablecaption{Orbital Parameters \label{orbittable}}
\tabletypesize{\normalsize}
\tablehead{ 
\colhead{Parameter} & \colhead{Value} 
}
\startdata
$P$ [days]                & 93.76  $\pm$  0.02 \\
$T_{\rm SC}$ [HJD]        & 2456641.66 $\pm$ 0.32 \\
$e$                       & 0.00 \\
$\gamma$ [{\kms}]         & 29.77 $\pm$ 0.64 \\
$K_{1}$ [{\kms}]          & 10.74 $\pm$ 1.17 \\
$K_{2}$ [{\kms}]          & 77.70 $\pm$ 1.22 \\
$\chi^{2}$                & 67.97 \\
$rms_{1}$                 & 4.50 \\
$rms_{2}$                 & 6.05 \\
$N_{spectra}$ (primary)       & 52 \\
$N_{spectra}$ (secondary)     & 49 \\
\hline
$q=M_{2}/M_{1}$           & 0.14 $\pm$ 0.02 \\
$M_1\sin ^3i$ [$M_\odot$] & 5.90 $\pm$ 0.30 \\
$M_2\sin ^3i$ [$M_\odot$] & 0.82 $\pm$ 0.11 \\
$a_1\sin i$ [AU]          & 0.09 $\pm$ 0.01 \\
$a_2\sin i$ [AU]          & 0.67 $\pm$ 0.01 \\
$a  \sin i$ [AU]          & 0.76 $\pm$ 0.01 \\
\enddata
\tablecomments{$T_{\rm SC}$ gives an epoch of superior conjunction.}
\end{deluxetable}

\section{Orbital Solution} \label{orbit}
To determine the orbital parameters of the HD~55606 system, we used the IDL code $rvfit$ \citep{marzoa15}. This program takes advantage of an Adaptive Simulated Annealing (ASA) global minimization routine to quickly determine the Keplerian orbital parameters given {\rv} data pertaining to one or more objects. In addition to a description of the code and associated error analysis, \citet{marzoa15} demonstrated consistency of the output with previous literature solutions for several examples of double-lined (SB2) and single-lined  (SB1) binaries as well as exoplanet systems. In the case of SB2s, the code solves for some or all of the orbital parameters including period ($P$), epoch of periastron passage ($T_{P}$), eccentricity ($e$), argument of periastron ($\omega$), systemic velocity ($\gamma$), and {\rv} semi-amplitudes of the primary ($K_{1}$) and secondary ($K_{2}$). One can fix or constrain each parameter in a simple input file. 

We began by allowing the orbital parameters to essentially vary freely, but due to a very small derived eccentricity being more than doubled by its error, we restricted the solution to a circular orbit ($e=0.0$). The resulting orbital parameters are given in Table~\ref{orbittable}. As $T_{P}$ is undefined for a circular orbit, ``$T_{\rm SC}$'' gives an epoch of superior conjunction of the Be star. Figure~\ref{orbitfig} shows the phased {\rv} data and associated residuals of the orbital solution. From this, it is clear that the development of the shell features occurs twice per orbit around conjunctures, and that the slope of the shell {\rv} generally corresponds to the motion of the Be star rather than the companion. One possible explanation of this behavior is that the shell features are formed in overdensities in the Be disk.

In addition to the SB2 orbital solution, we also obtained separate SB1 orbital solutions of both binary components. The SB1 solutions are not shown here since they yield the same circular orbit and $\sim93.8$ day period as the SB2 solution, with the {\rv} semi-amplitudes agreeing with the SB2 solution to within 0.7~{\kms}. The only significant discrepancy between SB1 solutions for the Be disk versus the sdOB disk is a $\sim6$ {\kms} difference in systemic velocities, whereby $\gamma=27$ {\kms} in the case of Be star SB1 while $\gamma=33$ {\kms} in the case of companion SB1. The SB2 solution hence prefers the average of the two, and this is reflected visually in the lower panel of Figure~\ref{orbitfig} by the vertical offsets of points away from $y=0$ and numerically in the high $\chi^{2}$ value reported in Table~\ref{orbittable} ($\chi^2=7.6$/35.3 for the Be/sdOB SB1 solutions).

The phased {\rv} measurements of the Be disk, sdOB disk, and transient shell components are given in Table~\ref{rvdatatable}, along with the number of spectral lines used to derive each {\rv} value. The equivalent width of the H$\eta$~3835~{\AA} shell component is also provided, to demonstrate the variable strength of the shell features while present, along with the peak separation of {\feii}~5317~{\AA}, which most clearly demonstrates the peak separation variability. The uncertainties on these quantities are negligible. 

\subsection{Inclination Angle}
Given the lack of permanent shell features in the HD~55606 spectra, the presence of shell features in most of the spectra, and the lack of any evidence of eclipses in an optical lightcurve despite detection of a signal corresponding to almost exactly half the orbital period (Labadie-Bartz et al., in preparation), the inclination angle of the system must be quite high though perhaps not exactly edge-on. 

A mass ratio of $M_{2}/M_{1}\sim0.14$ is indicated by the SB2 orbital solution, similar to what has been found for the previously known Be+sdO binaries. Assuming a lower limit of $i>75^{\circ}$ such that circumstellar disk gas can occasionally be seen projected against the Be star, thus producing shell lines \citep{hanuschik96}, the spectroscopic orbit suggests $M_{1}<6.6$~{\msun} for the Be star and $M_{2}<0.91$~{\msun} for the hot companion, leading to an upper limit on the total mass of $M_{\rm tot}<7.5$~{\msun}. This is comparable to the lower limit found by \citet{peters13} on the total mass of the known Be+sdO binary 59~Cyg ($M_{1}\sim6.3$~{\msun}, $M_{2}\sim0.6$~{\msun}). However, the total mass estimates known Be+sdO binaries -- $\phi$~Per, FY~CMa, HR~2142, and 60~Cyg -- are estimated to be $>10$~{\msun} \citep{mourard15,peters08,peters16,wang17}, placing HD~55606 toward the low mass end of this regime.

\begin{deluxetable*}{lccccccccccl}
\tablecaption{Radial Velocity Data \label{rvdatatable}}
\tabletypesize{\scriptsize}
\tablehead{ 
\colhead{HJD} & \colhead{Phase} & \colhead{$V_{r}$}  & \colhead{$N_{\rm lines}$} & \colhead{$V_{r}$}   & \colhead{$N_{\rm lines}$} & \colhead{$V_{r}$}  & \colhead{$N_{\rm lines}$} & \colhead{$W_{\lambda}$} & \colhead{{\vp}}        & \colhead{Instrument} \\
              &                 & \colhead{Be Disk}  & \colhead{Be Disk}         & \colhead{sdOB Disk} & \colhead{sdOB Disk}       & \colhead{Shell}    & \colhead{Shell}           & \colhead{H$\eta$ shell} & \colhead{{\feii} 5316} & \\
              &                 & \colhead{[{\kms}]} & \colhead{}                & \colhead{[{\kms}]}  & \colhead{}                & \colhead{[{\kms}]} & \colhead{}                & \colhead{[m{\AA}]}      & \colhead{[{\kms}]}     &   
}
\startdata
2458144.796 & 0.032 & 28.1 $\pm$ \hphantom{a}7.4 & 29 & \hphantom{a}49.7 $\pm$ \hphantom{a}9.3 & 4 & 39.0 $\pm$ 6.2 & 5 & 99 & 201 & ARCES \\
2456645.881 & 0.045 & 27.7 $\pm$ \hphantom{a}7.2 & 7 & \nodata & \nodata & \nodata & \nodata & \nodata & \nodata & APOGEE \\
2458053.964 & 0.063 & 32.0 $\pm$ 15.3 & 17 & \hphantom{a}61.1 $\pm$ \hphantom{a}7.7 & 3 & 59.4 $\pm$ 3.7 & 7 & 207 & 202 & SPM \\
2457491.796 & 0.067 & 23.3 $\pm$ 12.2 & 21 & \hphantom{a}65.6 $\pm$ \hphantom{a}9.7 & 5 & 50.7 $\pm$ 5.7 & 8 & 130 & 191 & FEROS \\
2458148.751 & 0.074 & 13.8 $\pm$ \hphantom{a}7.7 & 22 & \hphantom{a}58.7 $\pm$ \hphantom{a}4.4 & 2 & 39.1 $\pm$ 2.9 & 17 & 242 & 209 & ARCES \\
2458150.750 & 0.095 & 22.6 $\pm$ 10.3 & 25 & \hphantom{a}77.0 $\pm$ \hphantom{a}3.6 & 3 & 48.5 $\pm$ 2.8 & 17 & 337 & 209 & ARCES \\
2458151.795 & 0.107 & 22.7 $\pm$ \hphantom{a}9.6 & 25 & \hphantom{a}86.2 $\pm$ \hphantom{a}8.0 & 3 & 43.9 $\pm$ 2.6 & 18 & 336 & 221 & ARCES \\
2458154.792 & 0.139 & 20.7 $\pm$ 11.4 & 24 & 104.2 $\pm$ \hphantom{a}5.4 & 3 & 45.7 $\pm$ 2.6 & 14 & 281 & 216 & ARCES \\
2458157.797 & 0.171 & 18.7 $\pm$ \hphantom{a}9.1 & 27 & 105.9 $\pm$ \hphantom{a}7.2 & 4 & 44.9 $\pm$ 1.3 & 18 & 364 & 244 & ARCES \\
2458067.848 & 0.211 & 16.2 $\pm$ 11.5 & 29 & 110.3 $\pm$ \hphantom{a}8.9 & 3 & 38.8 $\pm$ 4.0 & 11 & 190 & 253 & ARCES \\
2458069.825 & 0.232 & 15.2 $\pm$ \hphantom{a}6.3 & 25 & 110.7 $\pm$ 17.1 & 3 & 38.0 $\pm$ 3.9 & 5 & 116 & 261 & ARCES \\
2456666.785 & 0.268 & 18.3 $\pm$ \hphantom{a}5.5 & 7 & \nodata & \nodata & \nodata & \nodata & \nodata & \nodata & APOGEE \\
2457792.864 & 0.278 & 16.8 $\pm$ \hphantom{a}6.5 & 26 & 108.6 $\pm$ \hphantom{a}6.2 & 2 & \nodata & \nodata & \nodata & 253 & ARCES \\
2457795.723 & 0.309 & 17.8 $\pm$ \hphantom{a}4.1 & 28 & 101.5 $\pm$ \hphantom{a}2.9 & 4 & \nodata & \nodata & \nodata & 251 & ARCES \\
2457796.688 & 0.319 & 17.7 $\pm$ \hphantom{a}3.9 & 29 & 100.1 $\pm$ \hphantom{a}4.4 & 2 & \nodata & \nodata & \nodata & 249 & ARCES \\
2458171.767 & 0.320 & 19.7 $\pm$ \hphantom{a}4.7 & 29 & 108.8 $\pm$ \hphantom{a}6.0 & 2 & \nodata & \nodata & \nodata & 254 & ARCES \\
2456671.771 & 0.321 & 17.6 $\pm$ \hphantom{a}7.1 & 7 & \nodata & \nodata & \nodata & \nodata & \nodata & \nodata & APOGEE \\
2456671.779 & 0.321 & 16.1 $\pm$ \hphantom{a}4.7 & 30 & \hphantom{a}94.5 $\pm$ \hphantom{a}7.8 & 4 & \nodata & \nodata & \nodata & 252 & ARCES \\
2452265.537 & 0.326 & 17.8 $\pm$ \hphantom{a}9.0 & 9 & \hphantom{a}99.8 $\pm$ 10.0 & 1 & \nodata & \nodata & \nodata & 247 & ELODIE \\
2458174.790 & 0.352 & 18.5 $\pm$ \hphantom{a}4.9 & 29 & 102.5 $\pm$ \hphantom{a}6.7 & 3 & \nodata & \nodata & \nodata & 245 & ARCES \\
2458176.657 & 0.372 & 17.4 $\pm$ \hphantom{a}4.5 & 29 & \hphantom{a}88.5 $\pm$ \hphantom{a}6.4 & 5 & \nodata & \nodata & \nodata & 243 & ARCES \\
2458083.849 & 0.382 & 16.4 $\pm$ \hphantom{a}6.0 & 30 & \hphantom{a}90.7 $\pm$ \hphantom{a}5.4 & 3 & \nodata & \nodata & \nodata & 244 & ARCES \\
2458179.769 & 0.405 & 21.4 $\pm$ \hphantom{a}6.4 & 29 & \hphantom{a}80.1 $\pm$ \hphantom{a}8.9 & 4 & \nodata & \nodata & \nodata & 239 & ARCES \\
2458086.816 & 0.414 & 19.7 $\pm$ \hphantom{a}5.2 & 30 & \hphantom{a}71.0 $\pm$ 11.1 & 5 & \nodata & \nodata & \nodata & 236 & ARCES \\
2458089.788 & 0.445 & 20.0 $\pm$ \hphantom{a}9.5 & 25 & \hphantom{a}50.9 $\pm$ \hphantom{a}4.7 & 3 & \nodata & \nodata & \nodata & 226 & ARCES \\
2458091.813 & 0.467 & 21.4 $\pm$ \hphantom{a}5.4 & 27 & \hphantom{a}42.5 $\pm$ \hphantom{a}5.5 & 5 & \nodata & \nodata & \nodata & 223 & ARCES \\
2458095.037 & 0.501 & 22.0 $\pm$ 14.7 & 16 & \hphantom{a}22.9 $\pm$ \hphantom{a}9.6 & 3 & \nodata & \nodata & \nodata & 215 & SPM \\
2458095.815 & 0.510 & 25.5 $\pm$ 13.9 & 27 & \hphantom{a}24.2 $\pm$ \hphantom{a}5.9 & 5 & \nodata & \nodata & \nodata & 203 & ARCES \\
2458097.056 & 0.523 & 24.9 $\pm$ 10.5 & 21 & \hphantom{a}22.1 $\pm$ 10.0 & 1 & \nodata & \nodata & \nodata & 202 & ARCES \\
2458097.791 & 0.531 & 28.8 $\pm$ 10.2 & 29 & \hphantom{a}22.4 $\pm$ \hphantom{a}4.2 & 2 & 17.7 $\pm$ 8.6 & 5 & 49 & 196 & ARCES \\
2452191.904 & 0.541 & 27.3 $\pm$ 10.4 & 21 & \hphantom{a}10.0 $\pm$ \hphantom{a}6.5 & 4 & \nodata & \nodata & \nodata & 195 & FEROS \\
2458098.826 & 0.542 & 23.8 $\pm$ 11.4 & 28 & \hphantom{a}14.8 $\pm$ \hphantom{a}7.3 & 4 & 18.8 $\pm$ 4.8 & 7 & 73 & 203 & ARCES \\
2458099.805 & 0.552 & 29.6 $\pm$ \hphantom{a}8.7 & 28 & \hphantom{a}14.7 $\pm$ 13.8 & 5 & 29.0 $\pm$ 3.4 & 11 & 118 & 190 & ARCES \\
2458100.814 & 0.563 & 28.9 $\pm$ \hphantom{a}7.8 & 29 & \hphantom{a}11.3 $\pm$ \hphantom{a}8.2 & 5 & 38.1 $\pm$ 9.2 & 15 & 277 & 193 & ARCES \\
2458101.805 & 0.573 & 29.2 $\pm$ 10.9 & 28 & \hphantom{a}\hphantom{a}1.3 $\pm$ \hphantom{a}9.3 & 2 & 43.4 $\pm$ 9.1 & 16 & 335 & 198 & ARCES \\
2457821.670 & 0.586 & 32.3 $\pm$ 13.3 & 27 & \hphantom{a}-4.0 $\pm$ \hphantom{a}4.8 & 4 & 51.7 $\pm$ 6.0 & 17 & 320 & 203 & ARCES \\
2457824.747 & 0.618 & 38.1 $\pm$ \hphantom{a}7.5 & 25 & -11.8 $\pm$ \hphantom{a}8.0 & 6 & 59.4 $\pm$ 3.7 & 15 & 296 & 220 & ARCES \\
2458111.813 & 0.680 & 33.5 $\pm$ 10.9 & 27 & -38.3 $\pm$ \hphantom{a}6.4 & 4 & 70.8 $\pm$ 6.3 & 11 & 224 & 242 & ARCES \\
2458112.864 & 0.691 & 35.5 $\pm$ 10.6 & 28 & -43.6 $\pm$ \hphantom{a}5.7 & 4 & 75.8 $\pm$ 6.6 & 10 & 140 & 247 & ARCES \\
2458113.816 & 0.702 & 35.7 $\pm$ \hphantom{a}5.1 & 28 & -45.4 $\pm$ \hphantom{a}6.8 & 4 & 75.9 $\pm$ 7.0 & 11 & 137 & 249 & ARCES \\
2458116.811 & 0.733 & 34.0 $\pm$ \hphantom{a}6.8 & 30 & -49.2 $\pm$ \hphantom{a}5.7 & 5 & 78.7 $\pm$ 8.1 & 12 & 187 & 252 & ARCES \\
2458117.817 & 0.744 & 34.9 $\pm$ \hphantom{a}6.2 & 30 & -49.9 $\pm$ \hphantom{a}6.6 & 5 & 78.3 $\pm$ 7.1 & 9 & 138 & 253 & ARCES \\
2458118.796 & 0.755 & 34.7 $\pm$ \hphantom{a}5.3 & 30 & -50.1 $\pm$ \hphantom{a}9.9 & 5 & 77.5 $\pm$ 7.6 & 10 & 168 & 255 & ARCES \\
2458119.791 & 0.765 & 35.6 $\pm$ \hphantom{a}4.4 & 29 & -47.6 $\pm$ \hphantom{a}6.2 & 4 & 75.9 $\pm$ 6.2 & 6 & 142 & 257 & ARCES \\
2458120.739 & 0.775 & 34.8 $\pm$ \hphantom{a}6.5 & 29 & -45.7 $\pm$ 12.9 & 5 & 80.8 $\pm$ 5.0 & 4 & 57 & 257 & ARCES \\
2457746.795 & 0.787 & 35.7 $\pm$ \hphantom{a}7.6 & 30 & -45.3 $\pm$ \hphantom{a}8.5 & 4 & 94.1 $\pm$ 4.5 & 9 & 110 & 253 & ARCES \\
2458123.780 & 0.808 & 38.4 $\pm$ \hphantom{a}4.1 & 29 & -35.7 $\pm$ \hphantom{a}6.4 & 4 & 85.0 $\pm$ 4.5 & 3 & 45 & 250 & ARCES \\
2457473.584 & 0.873 & 34.3 $\pm$ \hphantom{a}6.0 & 30 & -19.1 $\pm$ \hphantom{a}5.5 & 5 & \nodata & \nodata & \nodata & 244 & ARCES \\
2458130.812 & 0.883 & 43.4 $\pm$ \hphantom{a}5.2 & 29 & \hphantom{a}-8.6 $\pm$ \hphantom{a}4.8 & 3 & \nodata & \nodata & \nodata & 240 & ARCES \\
2457849.629 & 0.884 & 34.0 $\pm$ \hphantom{a}6.1 & 30 & -18.9 $\pm$ 14.8 & 6 & \nodata & \nodata & \nodata & 238 & ARCES \\
2458134.813 & 0.925 & 35.8 $\pm$ \hphantom{a}8.9 & 29 & \hphantom{a}\hphantom{a}2.4 $\pm$ \hphantom{a}6.0 & 2 & \nodata & \nodata & \nodata & 231 & ARCES \\
2458137.796 & 0.957 & 30.5 $\pm$ \hphantom{a}5.2 & 30 & \hphantom{a}\hphantom{a}9.1 $\pm$ \hphantom{a}5.3 & 3 & 36.0 $\pm$ 8.3 & 6 & 75 & 221 & ARCES \\
\enddata
\end{deluxetable*}

\begin{figure*}[h!]
\epsscale{1.0}
\plotone{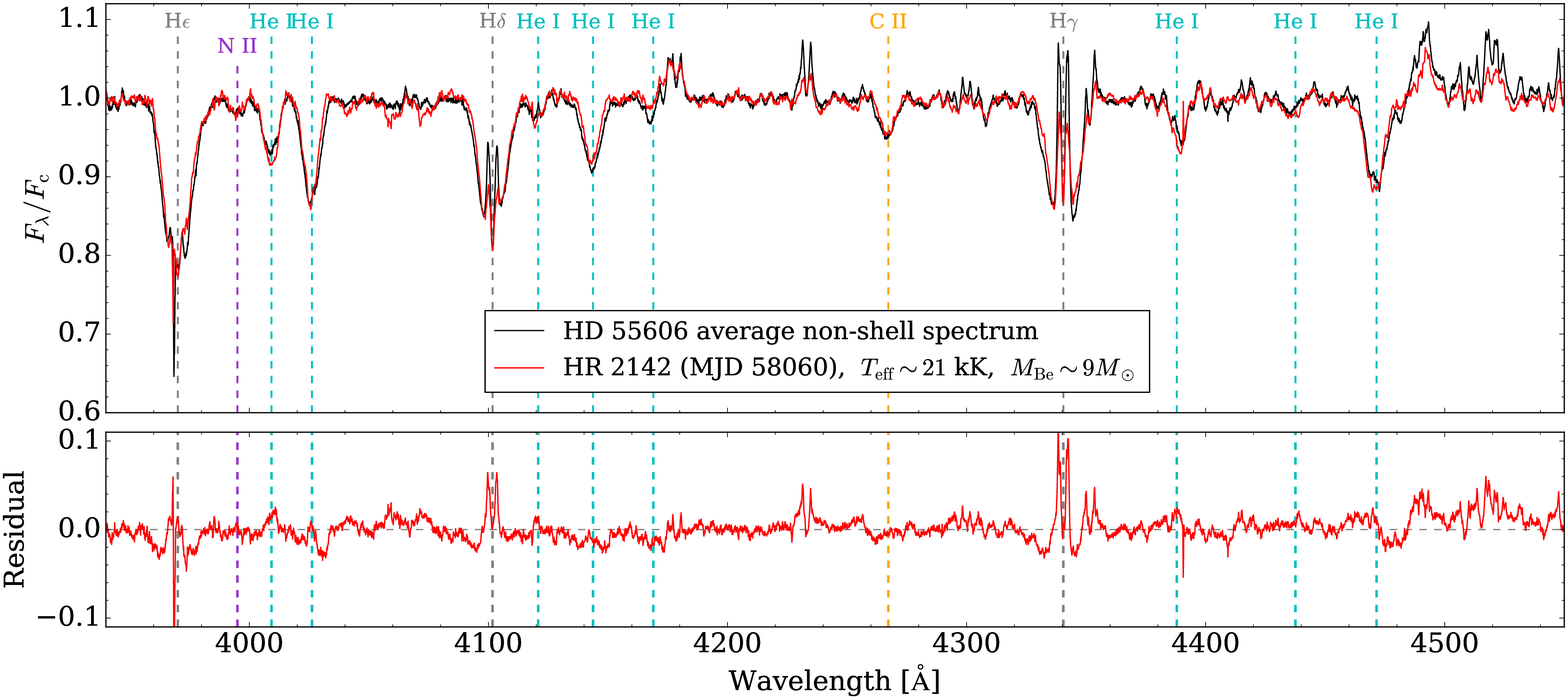}
\caption{Comparison of the HD~55606 average non-shell spectrum (black line) to a spectrum of HR~2142 (red line), with the lower panel showing residuals of HD~55606 minus HR~2142. Relevant absorption features are labeled. \label{hr2142fig}}
\end{figure*}

\section{Stellar Parameters} \label{stepars}
\subsection{Be Star Parameters}
There is a discrepancy between the Be star mass suggested by the orbital solution versus that suggested by our spectral classification and especially versus past spectroscopic analysis. As noted in Section~\ref{spectype}, past studies \citep{levenhagen06,fremat06} found $M_{\rm Be}>10$ {\msun} based on comparison to synthetic spectra, and our qualitative analysis of the spectra indicates a temperature class of B2, for which the associated Be star mass should be closer to 8~{\msun} \citep{torres10}. There are two possible explanations for this discrepancy, with the first being severe underestimation of the orbital motion of the Be star. For example, if $K_{\rm 1}=23.0$ {\kms} rather than 10.7 {\kms}, then the case of $i=77^{\circ}$ results in $M_{\rm 1}\sim8.0$ {\msun} and $M_{\rm 2}\sim2.4$ {\msun}. In this case, the companion would be the most massive subdwarf known in a binary with a Be star. Despite the {\rv} of both components of HD~55606 having been determined indirectly, using lines formed in the disks rather than on the stellar surfaces, we can find no way to increase $K_{\rm 1}$ and therefore we discard the possibility. The alternative explanation is that, similar to the sdOB binary companion, the HD~55606 Be star is hotter than expected for its mass. 

In Figure~\ref{hr2142fig}, we compare an average, non-shell HD~55606 spectrum to a spectrum of the previously known Be+sdO binary HR~2142 \citep{peters16} over the wavelength range covering the strong absorption features present for both stars. The HR~2142 spectrum was obtained on 2017 November 3 using the APO 3.5m telescope and ARCES spectrograph, and was handled identically to the ARCES spectra of HD~55606 in terms of data reduction, continuum flattening, and merging of orders. 

HD~55606 produces a slightly stronger emission spectrum compared to HR~2142, but the two stars are otherwise nearly spectroscopic twins based on the $>100$ Be stars we have observed with the ARCES instrument. Both stars exhibit \emph{exactly} the same range of emission and absorption species, with the absorption line strengths and widths being particularly comparable. The majority of differences between the spectra can be attributed to uncertainty in continuum level due to numerous weak emission features. For these reasons, we adopt for the HD~55606 Be star the effective temperature assumed for the HR~2142 Be star by \citet{peters16}, namely {\teff}~$=21$~kK. 

As for the Be star rotational velocity, the {\vsini}~$=350$~{\kms} measured for HD~55606 by \citet{levenhagen06} is a reasonable lower limit. Analysis of the broad Be star absorption lines using the IDL code \emph{IACOB Broad v7} \citep{simondiaz14} suggests a rotational velocity closer to 400~{\kms}, and up to 430~{\kms} in the case of {\hei}~4026~{\AA}. However, due to the questionably large macroturbulent velocity measurements ($v_{mac}>300$ {\kms}) and the excessive uncertainties on all output quantities, the results are not presented here. On average, the FWHM of {\hei}~4026~{\AA} is $\sim615$~{\kms}, such that HD~55606 is undoubtedly a very rapidly rotating Be star.

In comparison to the well-studied sample of eclipsing binary stars presented by \citet{torres10}, the HD~55606 Be star has a mass comparable to the B3V and B2.5V primary stars of V539~Ara ($M=6.24$~{\msun}, $R=4.52$~{\rsun}) and CV~Vel ($M=6.09$~{\msun}, $R=4.09$~{\rsun}). Both of these stars are relatively slow rotators, with the V539~Ara primary having {\vsini}~$=75$~{\kms} and the CV~Vel primary having {\vsini}~$=19$~{\kms}, such that rotational flattening is not a significant concern. Since the HD~55606 Be star is likely a near critical rotator however, the ratio of equatorial to polar radii should be $R_{\rm E}/R_{\rm P}\approx1.5$ \citep{zahn10}. The average radius of the V539~Ara and CV~Vel primaries is $R=4.30$~{\rsun}, and if we assume this is the radius of a non-rotating version of the HD~55606 primary, then the critically rotating version has $R_{\rm E}\approx5.8$~{\rsun}, $R_{\rm P}\approx3.8$~{\rsun}, and a mean radius of $R_{0}\approx5.1$~{\rsun} \citep[see Table~1 of ][]{zahn10}.

Under the assumptions of {\teff}~$=21$~kK and $R_{0}=5$~{\rsun}, the HD~55606 Be star may indeed be overly hot and hence overly luminous for a star of $M\sim6.2$~{\msun}. This conclusion was also reached by \citet{gies98} in the case of the Be+sdO binary $\phi$~Per, but lack of a reliable parallax estimate prevents us from confirming that it applies to HD~55606 as well.

\subsection{sdOB Star Parameters}
Although UV spectra of HD~55606 are unfortunately not available to try and directly detect the sdOB star's spectrum, the presence of {\hei} emission and {\heii} absorption indicates that the sdOB star has an effective temperature of {\teff}$>27$ kK. As for the rotational velocity, the FWHM~$\sim160$~{\kms} of the average {\heii}~4686~{\AA} absorption line (see Figure~\ref{heidemo}) can be used to set an upper limit of {\vsini}~$<80$~{\kms} for the sdOB star.

\begin{deluxetable}{lcc}
\tablecaption{Basic parameters of the Be and sdOB components of HD~55606. \label{newparmtable}}
\tabletypesize{\normalsize}
\tablehead{ 
\colhead{Parameter} & \colhead{Be star}   & \colhead{sdOB star}
}
\startdata
$M$ [{\msun}]             & \textbf{5.97 -- 6.55}   & \textbf{0.83 -- 0.90} \\
$R$ [{\rsun}]             & $\mathbf{\sim5}$      & \nodata               \\
{\teff} [kK]              & $\mathbf{\sim21}$     & $\mathbf{>27}$        \\
{\vsini} [{\kms}]         & $\mathbf{>350}$       & $\mathbf{<80}$        \\
$L$ [{\lsun}]             & 4360                  & \nodata               \\
log$\,g$ [dex]            & 3.82 -- 3.86          & \nodata               \\
{\vcrit} [{\kms}]         & $>390$                & \nodata               \\
$W=${\vrot}/{\vcrit}      & $>0.89$               & \nodata               \\
\hline
$M_{\rm tot}$ [{\msun}]   & \multicolumn{2}{c}{\textbf{6.80 -- 7.45}}     \\
\enddata
\tablecomments{The mass ranges pertain to an inclination angle range of $i=75-85^{\circ}$. Measured, estimated, and assumed quantities are given in bold font to distinguish them from derived quantities.}
\end{deluxetable}

Table~\ref{newparmtable} gives estimates of the stellar parameters of the HD~55606 binary components. For the Be star, derived estimates are provided for luminosity ($L$), surface gravity (log$\,g$), critical rotation velocity ({\vcrit}), and critical fraction ($W$).

\begin{figure*}[t!]
\epsscale{1.0}
\plotone{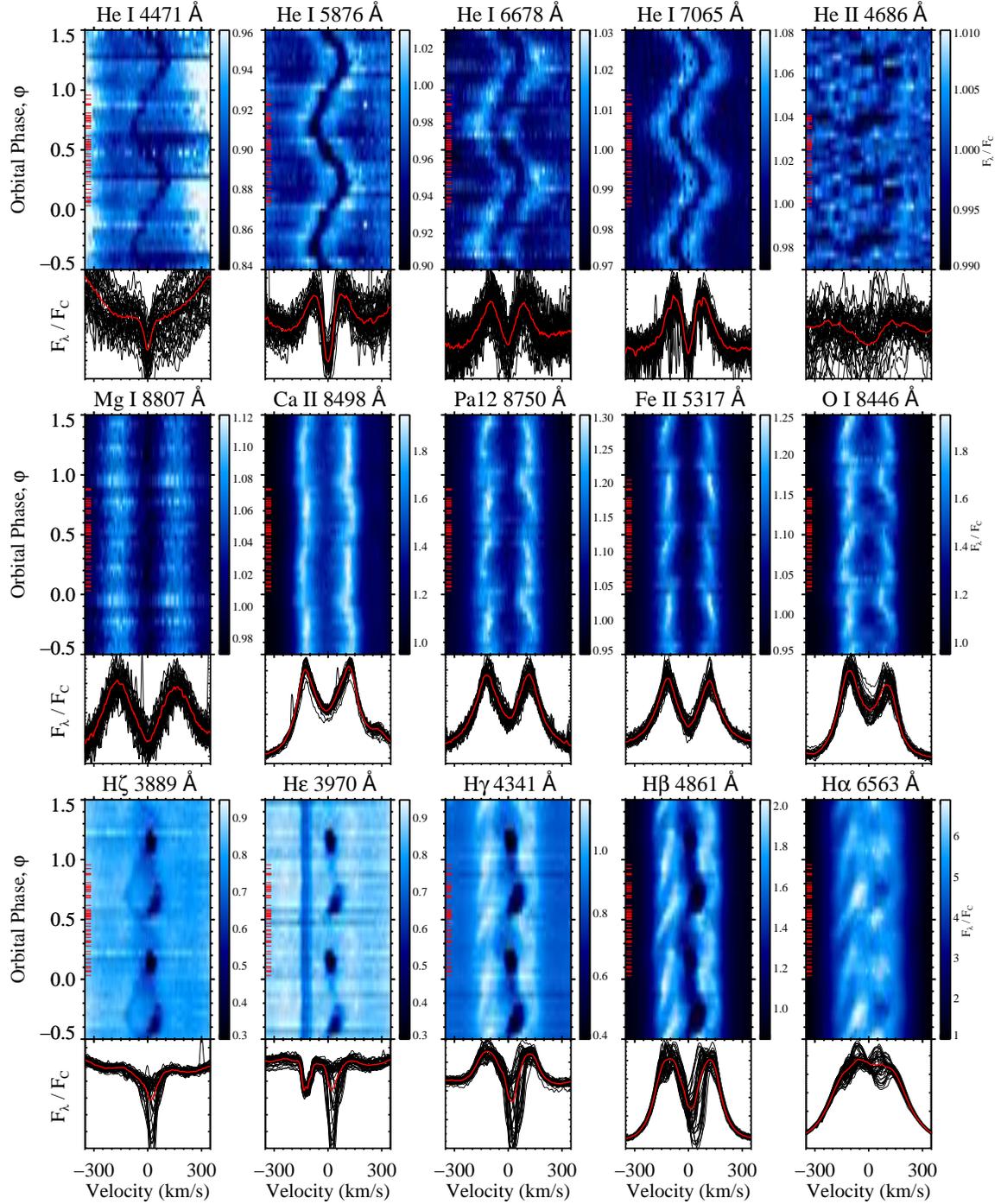}
\caption{Interpolated orbital phase diagrams and profiles of various spectral lines. In the phase diagrams, the $\gamma$ value given in Table~\ref{orbittable} has been subtracted, but otherwise no {\rv} correction has been applied. The orbit has been reflected by $\phi=\pm0.5$ with red ticks at left indicating the observations and with colors bars indicating the height relative to continuum level. In the line profile panels, black lines are the individual {\rv}-corrected spectra and the flux y-range is identical to that given by the color bars above. Thick red lines are the mean line profiles, in the rest frame of the sdOB star for {\hei}/{\heii} and otherwise in the rest frame of the Be star. The top row shows {\hei} lines in which the sdOB star is most clearly detected as well as {\heii}~4686~{\AA}, where very weak absorption can clearly be seen at velocities corresponding to the sdOB star orbit. The middle row shows representative Be disk metallic emission lines and Pa12, highlighting the phase-locked {\vp} variability seen in most lines. Also note the significantly larger and less variable {\vp} of the {\mgi}~8806~{\AA} line. The bottom row shows five Balmer series lines, with the H$\zeta$ and H$\epsilon$ panels zoomed in the cores of the broad absorption to focus on the transient shell features. Note that in the H$\zeta$ panel, the orbit of the sdOB star can be seen in the {\hei}~3888~{\AA} line, and that in the H$\epsilon$ panel, the vertical features are composite interstellar absorption in {\caii} H (3968~{\AA}). \label{dynamMontage}}
\end{figure*}

\begin{figure}[htb!]
\epsscale{1.0}
\plotone{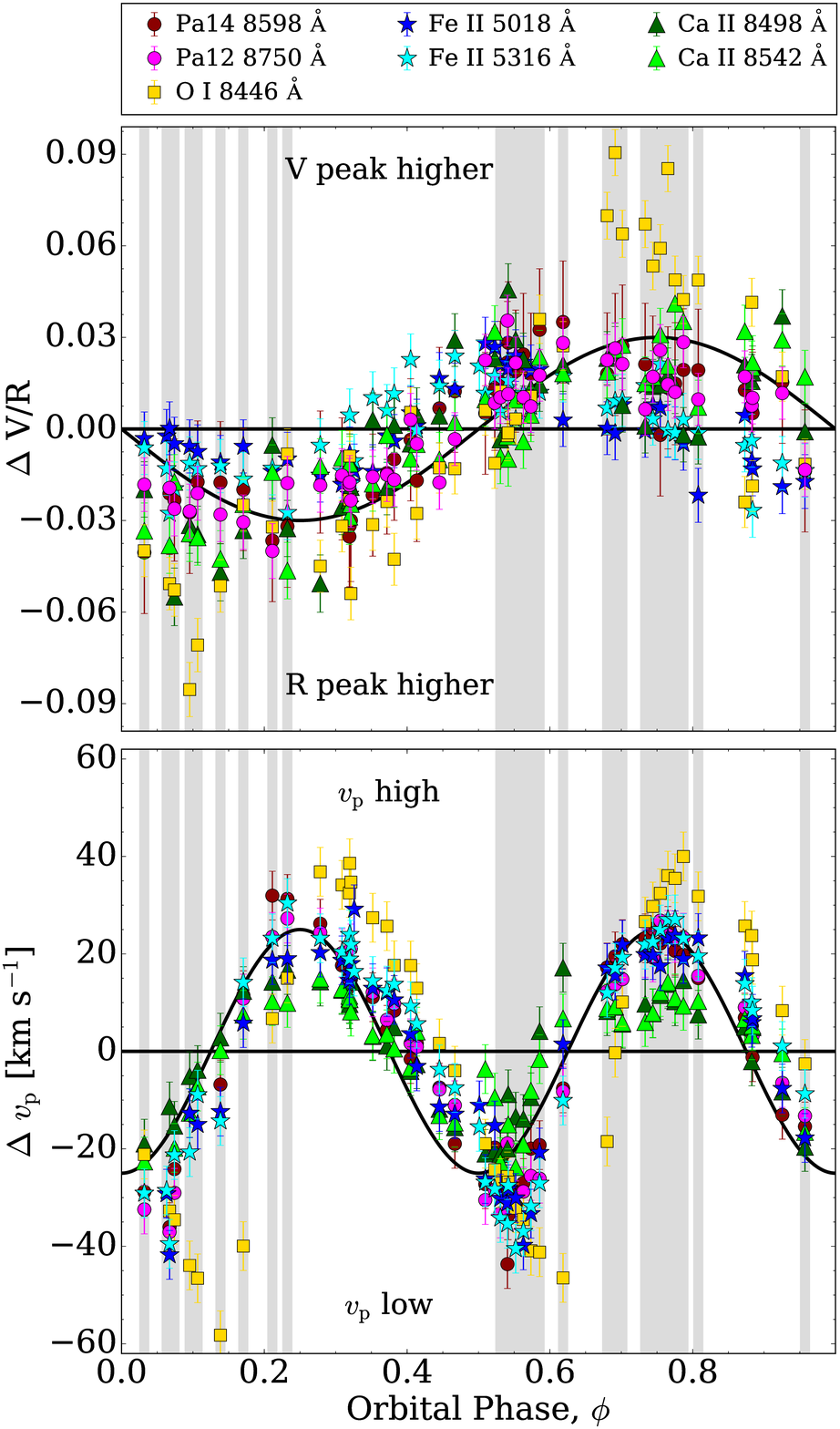}
\caption{Double-peak intensity ratios (V/R) and peak separations ({\vp}) of representative strong lines, phased to the orbital period. For each line, the average V/R and {\vp} have been subtracted from the multi-epoch values. The grey vertical stripes indicate epochs where the transient shell components are present. \label{phaseplot}}
\end{figure}

\section{Orbital Variations} \label{phaselocked}
\subsection{Helium} 
The upper row of Figure~\ref{dynamMontage} displays phased dynamical spectra of the strong {\hei} lines in which the sdO star orbit can be clearly seen, as well of {\heii}~4686~{\AA}, which is very weakly detected. {\hei} emission arising in the sdOB disk is most obvious in {\hei}~5876, 6678, and 7065~{\AA}, whereas for {\hei}~4471, 4026, and 4713~{\AA}, the sdOB signature is a narrow absorption embedded in the broad absorption cores of the Be star. The {\hei}~6678~{\AA} line shows some evidence of phase-locked V/R variability, with the R peak being bright at $\phi=0.25$ and the V peak being bright at $\phi=0.75$. This behavior is not repeated in the other {\hei} lines however.

\subsection{Metals \& Paschen Series} 
The middle row of Figure~\ref{dynamMontage} shows similar phase diagrams of representative metallic emission lines formed in the Be disk. Pa12~8750~{\AA} is included since the variability of the Paschen series lines is identical to that of the metallic lines. Although the orbital variation of the Be star is almost too negligible to be seen here, it is clear that the {\vp} of most lines vary in a complicated fashion that is locked to the orbital phase. Namely, {\vp} gradually decreases before conjunctures, before rather suddenly increasing up to the maximum values around quadratures. This is most obvious in the cases of {\feii}~5317~{\AA} and {\oi}~8446~{\AA}, but the pattern can also be seen in the {\caii} triplet and Paschen series lines. 

On the other hand, many of the neutral metal lines (e.g. {\mgi}~5183, {\mgi}~8806~{\AA}, {\ci}~16895~{\AA}) are immune to the {\vp} variability. These lines exhibit {\vp} that are up to $\sim100$ {\kms} wider than those of the {\feii} and {\feii}-like lines, indicating that they are inner disk lines and hence may be shielded from interactions between the sdOB star and the Be disk that are presumably responsible for the variability of the {\feii} and {\feii}-like lines. 

In the majority of Be+sdO binaries, the primary signature of the sdO star is single-peaked or highly V/R-variable emission in the {\hei} lines \citep{stefl00,maintz05,peters13}, with the single or stronger peak corresponding to the side of the Be disk facing the hot companion and hence being heated/irradiated. This is not the case for HD 55606, in which the {\hei} emission is formed in a separate disk around the sdOB star. Nonetheless, many of the Be disk emission lines of HD~55606 do in fact exhibit orbital-phase-locked V/R variability that may be caused by heating of the side of the disk facing the sdOB star. 

\subsection{Balmer Series} 
As demonstrated in the lower row of Figure~\ref{dynamMontage}, and particularly in the {\hb} panel, the Balmer series lines exhibit a complex form of phase-locked variability. The V peak appears to be composed of two intertwined/braided peaks, with the outer (most negative velocity) peak reaching maximum intensity twice per orbit around $\phi\sim0.4$ and $\phi\sim0.9$. The R peak, on the other hand, is less of a peak and more of a plateau, interrupted twice per orbit by the occurrence of transient shell phases. The transient shell contributions are most obvious in the H$\epsilon$ and H$\zeta$ panels of Figure~\ref{dynamMontage}, and as has been already discussed in relation to Figure~\ref{orbitfig}, the shell phases occur twice per orbit and are present for about half of each orbit. The shell phase starting around inferior conjunction begins at a velocity slightly lower than that of the Be disk (and corresponding to the {\rv} of the sdOB star at that point), with the velocity gradually increasing toward quadrature. The maximum observed shell {\rv} of 94 {\kms} occurred in our data set at $\phi=0.79$, but the shell phase observed starting around superior conjunction takes place at lower (35--60 {\kms}) and less variable velocities. 

\subsection{Phase-locked V/R and Peak Separation Variability}  \label{phaselock}
Figure~\ref{phaseplot} provides a quantitative demonstration of phase-locked variability in the HD 55606 spectra. The two panels plot the residuals of the Be disk line V/R ratios and peak separations, phased to the $\sim$93.8-day orbital period. Although the trends in V/R and {\vp} are quite obvious in Figure~\ref{phaseplot}, we proceeded to perform orbit-independent period searches of the V/R and {\vp} measurements for numerous metallic and Paschen series lines by supplying the $rvfit$ code (see Section~\ref{orbit}) with V/R and {\vp} rather than {\rv}. The results of several representative lines are presented in Tables~\ref{vrRatioTable} and \ref{peaksepPeriodTable}, which give the variability periods, the mean V/R or {\vp} values, the variability semi-amplitudes, the ratios of the variability periods over the orbital period, and the phase shift with respect to the orbital ephemeris. In terms of amplitude of {\vp} and V/R variability, the overall most variable metallic line in the HD~55606 spectra is {\oi}~8446~{\AA}.

\begin{deluxetable}{lccccr}
\caption{V/R variability periods \label{vrRatioTable}}
\tabletypesize{\scriptsize}
\tablehead{ 
\colhead{Line} & \colhead{$P$}    & \colhead{$<$V/R$>$} & \colhead{$K$} & \colhead{$P$/$P_{\rm orb}$} & \colhead{$\delta\phi$} \\
\colhead{}     & \colhead{[days]} & \colhead{}          & \colhead{}    & \colhead{}                    & \colhead{}
}
\startdata
Pa14 8598 {\AA} & 92.63 & 1.00 & 0.03 & 0.99 & 0.10 \\
Pa12 8750 {\AA} & 93.94 & 1.00 & 0.03 & 1.00 & -0.09 \\
{\feii} 5018 {\AA} & 92.15 & 1.00 & 0.01 & 0.98 & 0.07 \\
{\feii} 5317 {\AA} & 92.18 & 1.01 & 0.02 & 0.98 & 0.07 \\
{\oi} 8446 {\AA} & 93.66 & 1.06 & 0.06 & 1.00 & -0.01 \\
{\caii} 8498 {\AA} & 93.66 & 0.98 & 0.03 & 1.00 & -0.01 \\
{\caii} 8542 {\AA} & 91.76 & 0.97 & 0.03 & 0.98 & 0.22 \\
\hline
Mean & 92.85 & 1.00 & 0.03 & 0.99 & 0.05 \\
\enddata
\end{deluxetable}

\begin{deluxetable}{lccccr}
\caption{Peak Separation variability periods \label{peaksepPeriodTable}}
\tabletypesize{\scriptsize}
\tablehead{ 
\colhead{Line} & \colhead{$P$}    & \colhead{$<$Peak Sep.$>$} & \colhead{$K$}      & \colhead{$P$/$P_{\rm orb}$} & \colhead{$\delta\phi$} \\
\colhead{}     & \colhead{[days]} & \colhead{[{\kms}]}        & \colhead{[{\kms}]} & \colhead{}                    & \colhead{}
}
\startdata
Pa14 8598 {\AA} & 46.86 & 239 & 27 & 0.50 & -0.10 \\
Pa12 8750 {\AA} & 46.92 & 234 & 27 & 0.50 & -0.11 \\
{\feii} 5018 {\AA} & 46.86 & 231 & 26 & 0.50 & -0.09 \\
{\feii} 5317 {\AA} & 46.90 & 230 & 28 & 0.50 & -0.09 \\
{\oi} 8446 {\AA} & 46.96 & 208 & 39 & 0.50 & -0.08 \\
{\caii} 8498 {\AA} & 46.85 & 238 & 15 & 0.50 & -0.12 \\
{\caii} 8542 {\AA} & 46.95 & 234 & 14 & 0.50 & -0.13 \\
\hline
Mean & 46.90 & 230 & 25 & 0.50 & -0.10 \\
\enddata
\end{deluxetable}

These results confirm that the V/R ratios are locked to the orbital period and that the {\vp} variability indeed repeats twice per orbit. The former result is expected in the case of the hot sdOB star illuminating the side of the Be disk directly exposed to it \citep[e.g.][]{stefl00}, but the latter result has not been previously noted in Be+sdO binaries.

\section{Discussion} \label{discussion}
Given the orbital and stellar parameters of HD~55606 in addition to the measured {\vp}, it is possible to estimate the Roche lobe and disk radii of the binary companions. If we assume that the emission peaks of both the Be disk and the sdOB disk sample Keplerian motion at some characteristic radii within the disks, then we can equate the half emission peak separations with the projected Keplerian velocities. The Keplerian circular velocity is

\begin{equation} \label{eq2}
v_c = \sqrt{{GM}\over{R}} = 436.8~{\rm km~s^{-1}}\sqrt{{M/M_\odot}\over{R/R_\odot}}
\end{equation}

The emission line peak separation is related to $v_{c}$ via 

\begin{equation} \label{eq3}
v_{\rm p} = 2\,v_c \sin i
\end{equation}

Solving for $v_{c}$ in equation~\ref{eq3} and plugging the result into equation~\ref{eq2} yields the following expression for an estimate of the disk radii.

\begin{equation} \label{eq4}
 R_{\rm disk}/R_\odot = \left(\frac{873.6~{\rm km~s^{-1}} \sin i}{v_{\rm p}}\right)^{2} M/M_\odot
\end{equation}

To check on whether Roche lobe overflow of the disks should be occurring, the mass ratio derived in Section~\ref{orbit} ($q=0.14$) can be fed into the simple analytical prescription for Roche lobe radius given by \citet{eggleton83}. 

\begin{deluxetable}{lcc}
\tablecaption{Roche radii and mean disk radii for an inclination angle range of $i=75-85^{\circ}$. \label{radiustable}}
\tabletypesize{\normalsize}
\tablehead{ 
\colhead{Parameter} & \colhead{[AU]}   & \colhead{[$R_{\rm Be}\approx5$ {\rsun}]}
}
\startdata
$R_{\rm Roche,\,Be}$                    & 0.424 -- 0.437  & 18.2 -- 18.8 \\
$R_{\rm disk,\,Be}$ ({\feii}/{\caii})   & 0.384 -- 0.396  & 16.5 -- 17.0 \\
$R_{\rm disk,\,Be}$ ({\oi}~8446~{\AA})  & 0.486 -- 0.501  & 20.9 -- 21.5 \\
$R_{\rm disk,\,Be}$ ({\mgi}~8806~{\AA}) & 0.195 -- 0.202  & 8.40 -- 8.67 \\
\hline
$R_{\rm Roche,\,sdOB}$                  & 0.138 -- 0.143  & 5.95 -- 6.13 \\
$R_{\rm disk,\,sdOB}$ ({\hei})          & 0.105 -- 0.108  & 4.51 -- 4.65 \\
\enddata
\end{deluxetable}

The Roche lobe and disk radius estimates are given in Table~\ref{radiustable}, as calculated for $i=75^{\circ}-85^{\circ}$ and provided in both astronomical units (AU) and in Be star radii (assuming $R_{\rm Be}=5$ {\rsun}). For the Be disk, radii were estimated using the average {\feii}/{\caii} {\vp} of 234 {\kms}, the average {\oi}~8446~{\AA} {\vp} of 208 {\kms}, and the average {\mgi}~8806~{\AA} {\vp} of 328 {\kms}. For the sdOB disk, the radius was estimated using the 167 {\kms} average {\vp} of {\hei}~5876, 6678, and 7065~{\AA}. We note that these indirect disk radii estimates are subject to a number of uncertainties regarding the density distributions of material in the disks, possible truncation of the Be star disk by the companion object, and possible existence of a circumbinary disk.

The results suggest that the Be star disk extends out to or beyond the Roche radius, such that mass transfer may be ongoing. This is especially true in the case of {\oi}~8446~{\AA}, for which the preferential emitting region extends $\sim15$\% beyond the projected Be star Roche lobe radius. The enhanced variability of the line with respect to other features is therefore probably a result of radiative or kinematic interaction of the sdOB star. The {\mgi}~8806~{\AA} line represents the other side of the coin, with the lack of any substantial variability being due to little or no interaction with the sdOB star.

\begin{figure}[htb!]
\epsscale{1.0}
\plotone{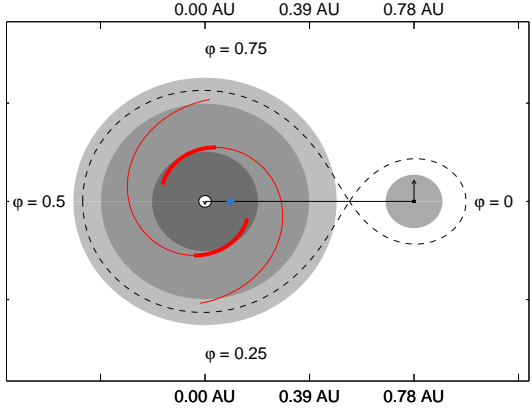}
\caption{Pole-on diagram of the HD~55606 system, showing the Be star on the left and the sdOB star on the right. Arrows above each star indicate the orbital motions of the stars, which are separated by $\sim0.78$ AU in the case of $i=80^{\circ}$, with the Be star located $\sim0.09$ AU from the center of mass (blue plus sign). Dashed curves show the Roche lobes associated with $M_2/M_1=0.14$, shaded regions show the circumstellar disks, and orbital phases are indicated. Differing shading of the Be disk indicates the preferential emitting radii of {\mgi}~8806~{\AA} (dark shading), {\feii}/{\caii} (medium shading), and {\oi}~8446~{\AA} (light shading). The red spiral arms indicate density perturbations in the Be disk, with the thicker red lines indicating the regions that give rise to the shell phases and to the peak separation variability. \label{diagramfig}}
\end{figure}

In Figure~\ref{diagramfig} we present a possible explanation for the transient shell phases and peak separation variability of HD~55606. The system is shown as viewed from a pole-on orientation, with physical dimensions as estimated in previous sections. Red curves on the Be disk show a two-armed spiral density perturbation based on the tidal shock formulation of \citet{ogilvie02}, in which the free parameter $\epsilon=c_{\rm s}\,{\rm sin}\,i/(K_{1}+K_{2})$ determines the radius of curvature function of the spiral arms. We adopted $i=80^{\circ}$ and the disk sound speed of $c_{\rm s}=14$~{\kms} assumed by \citet{peters16} in the case of HR~2142, resulting in $\epsilon=0.156$. Thicker parts of the red curves indicate overdensities near the Be star where gas can be viewed in projection against the stellar surface. Over the course of an orbit, these overdense regions intersect our line of sight to the Be star twice for a total time of about half an orbit, thus causing the observed shell phases. Small changes of $\pm0.02$ to the epsilon parameter cause the overdense regions to eclipse the Be star for significantly larger or smaller fractions of an orbit, such that our adopted value is consistent with observations. The differing radial velocity evolution of the $\phi=0$ shell phase versus the $\phi=0.5$ shell phase can be explained by the spiral structure of the overdensities. At $\phi=0.5$ we see the high density, inner arm approaching us at relatively high speed, while at $\phi=0$, the high density region occurs further out at a correspondingly lower velocity. During the shell phases, the increased emission peak separations are simply caused by enhanced line formation at higher velocities, closer to the star.

The complicated negative velocity portions of the strong Balmer series line profiles (see Figure~\ref{dynamMontage}) are also likely a natural result of the spiral density perturbations in the Be disk. Each time a high density region crosses our line of sight, the outer part of the spiral arm on the opposite side of the Be star gives rise to a bright peak at very negative {\rv} that gradually migrates to higher velocities until the arrival of the second high density region across our line of sight. At this point, the process repeats, leading to the apparent `braided' pattern of the V peak of the Balmer series lines, seen most clearly in {\hb}. Despite the line profiles of HD~55606 being considerably more complex that those produced by theoretical simulations of Be disks with binary-induced spiral arms, the basic notion of two V/R reversals per orbit is expected based on \citet{panoglou18}.

\section{Conclusions} \label{conclusions}
The star HD~55606 has been identified for the first time as an exotic binary system with a massive Be type primary star and a compact, hot companion that is probably an sdO or sdB star, bringing the total number of known or strongly suspected Be+sdOB binaries to twenty. These systems are the low mass analogues of Wolf-Rayet stars formed in interacting binaries \citep{gotberg18}, where past large-scale mass transfer resulted in the high temperature and low mass of the stripped-envelope secondary. In Be+sdOB binaries, the rapid rotation and ability to maintain a Be disk for the primary is also a result of mass transfer. This makes Be+sdOB binaries extremely interesting considering that the origin of rapid rotation and ability to form disks are unexplained in the vast majority of Be stars. It is possible that a significant fraction of Be stars attain these traits through binary interactions, but the companions are too faint and/or the orbital motions of the Be stars too small to permit easy detection \citep{wang18}. Indeed, HD~55606 is similar to the previously known Be+sdO binaries in terms of low $K_{1}$ and high inclination angle. 

Although the binary nature of HD~55606 was discovered somewhat serendipitously during an investigation of rare $H$-band spectral peculiarities, the optical spectra are remarkably similar to spectra of the previously known Be+sdO binary HR~2142, such that the commonalities may be useful toward a systematic search for similar binaries (e.g. using the BeSS database). Beyond the obvious smoking gun of radial velocity variable {\hei}/{\heii} emission/absorption, other spectral signatures include relative permanence of the Be star disk (i.e. no epochs when emission lines are not observed), triple-peaked or otherwise non-standard {\ha} line profiles, strong metallic emission (particularly {\feii} and the IR {\caii} triplet). These may simply be selection effects associated with very massive Be disks, but in lieu of a new UV spectroscopic survey of Be stars to search directly for stripped-envelope companions, it would be worthwhile to investigate Be stars exhibiting spectra similar to HD~55606 and HR~2142. 

\vspace{0.5cm}
\scriptsize{\emph{Acknowledgements.} Based on observations obtained with the Apache Point Observatory 3.5-meter telescope, which is owned and operated by the Astrophysical Research Consortium. Also based on observations obtained with the MPG-ESO 2.2m telescope at the European Southern Observatory (La Silla, Chile), under the programme 097.A-9024 and the agreement MPI-Observat\'orio Nacional-MCTIC (Brazil). This research has made use of the SIMBAD database, operated at CDS, Strasbourg, France.

J.P.W. was supported by NSF-AST 1412110. R.E.M. acknowledges support by VRID-Enlace 216.016.002-1.0 and the BASAL Centro de Astrof\'{\i}sica y Tecnolog\'{\i}as Afines (CATA) PFB--06/2007. D.G was supported by the National Science Foundation under Grant AST-1411654. G.S.S. was supported in part by NASA Gran NNX13AF34G. DP acknowledges financial support from Conselho Nacional de
Desenvolvimento Cient\'ifico e Tecnol\'ogico (Brazil) through grant 300235/2017-8.

The authors thank Matt Shultz for sharing the software used to make Figure~\ref{dynamMontage}.

\end{document}